\documentclass[aps,prl,citeautoscript,twocolumn,longbibliography,superscriptaddress]{revtex4-1}
\usepackage[T1]{fontenc}
\usepackage{latexsym}
\usepackage{graphicx} 
\usepackage{epstopdf}
\usepackage{amsmath}  
\usepackage{amssymb}
\usepackage{comment}
\usepackage{cases}
\usepackage{lipsum}
\usepackage{float}
\usepackage{tikz}
\usepackage{makecell}
\usepackage{bbold}

\usepackage[breaklinks=true]{hyperref}

\begin{document}

\title{A unified theory of spin and charge excitations in high-$T_c$ cuprates: Quantitative comparison with experiment and interpretation}
\author{Maciej Fidrysiak}
\email{maciej.fidrysiak@uj.edu.pl}
\affiliation{Institute of Theoretical Physics, Jagiellonian University, ul. {\L}ojasiewicza 11, 30-348 Krak{\'o}w, Poland }
\author{J{\'o}zef Spa{\l}ek}%
\email{jozef.spalek@uj.edu.pl}
\affiliation{Institute of Theoretical Physics, Jagiellonian University, ul. {\L}ojasiewicza 11, 30-348 Krak{\'o}w, Poland }

\begin{abstract}
  We provide a unified interpretation of both paramagnon and plasmon modes in high-$T_c$ copper-oxides, and verify it quantitatively against available resonant inelastic $x$-ray scattering (RIXS) data across the hole-doped phase diagram. Three-dimensional extended Hubbard model, with included long-range Coulomb interactions and doping-independent microscopic parameters for both classes of quantum fluctuations, is used. Collective modes are studied using VWF+$1/\mathcal{N}_f$ approach which extends variational wave function (VWF) scheme by means of an expansion in inverse number of fermionic flavors ($1/\mathcal{N}_f$). We show that intense paramagnons persist along the anti-nodal line from the underdoped to overdoped regime and undergo rapid overdamping in the nodal direction. Plasmons exhibit a three-dimensional character, with minimal energy corresponding to anti-phase oscillations on neighboring $\mathrm{CuO_2}$ planes. The theoretical spin- and charge excitation energies reproduce semi-quantitatively RIXS data for $\mathrm{(Bi, Pb)_2 (Sr, La)_2 CuO_{6+\delta}}$. The present VWF+$1/\mathcal{N}_f$ analysis of dynamics and former VWF results for static quantities combine into a consistent description of the principal properties of hole-doped high-$T_c$ cuprates as strongly correlated systems.
\end{abstract}

\maketitle

\emph{Introduction}---A profound problem in condensed matter physics is to unveil the microscopic structure of both the single- and many-particle excitations in high-temperature (high-$T_c$) cuprate superconductors (SC) as they evolve from antiferromagnetic (AF) insulator, through the SC phase, to a Fermi-Liquid normal state \cite{KeimerNature2015}. At low doping, localized holes coexist with collective spin-wave excitations which are now well understood within the framework of Heisenberg-type models \cite{ColdeaPhysRevLett2001,PengNatPhys2017}. Much less is known about the microscopic mechanism governing single- and many-particle excitations at moderate- and high doping levels, where no AF or charge order occur. Itinerant carriers are expected to cause Landau overdamping of spin-wave modes, particularly after AF order has been suppressed. On the contrary, resonant inelastic $x$-ray scattering (RIXS) and inelastic neutron scattering experiments demonstrate that robust \emph{paramagnons} persist across whole hole-doping phase diagram \cite{DeanNatMater2013,IshiiNatCommun2014,LeeNatPhys2014,GuariseNatCommun2014,WakimotoPhysRevB2015,MinolaPhysRevLett2017,IvashkoPhysRevB2017,MeyersPhysRevB2017,ChaixPhysRevB2018,Robarts_arXiV_2019,ZhouNatCommun2013,GretarssonPhysRevLett2016,FumagalliPhysRevB2019,LeTaconNatPhys2011,JiaNatCommun2014,PengPhysRevB2018}. In addition, RIXS provides evidence for low-energy charge modes (\emph{acoustic plasmons}) in both hole- and electron-doped cuprates \cite{IshiiPhysRevB2017,HeptingNature2018,IshiiJPhysSocJapan2019,LinNPJQuantMater2020,SinghArXiV2020,NagArxiv2020}. Over the years, several distinct high-$T_c$ SC mechanisms, based either on fluctuations (magnetic \cite{WakimotoPhysRevLett2004,DahmNatPhys2009,LeTaconNatPhys2011} and charge \cite{GrilliPhysRevLett1991,PeraliPhysRevB1996}), or local correlations \cite{SpalekPhysRevB2017}, have been proposed. In effect, a unified quantitative theory of the equilibrium thermodynamic properties, as well as correlated single-particle and collective excitations in high-$T_c$ copper-oxides, is now in demand to single-out the microscopic SC pairing scenario.

The current theoretical frameworks, used interpret RIXS data for copper-oxides, encompass determinant quantum Monte-Carlo (DQMC) \cite{PengPhysRevB2018},  Hubbard-operator large-$N$ limit \cite{GrecoPhysRevB2016,GrecoJPSJ2017,GrecoPhysRevB2020}, random-phase-approximation (RPA) \cite{GuariseNatCommun2014}, and spin-wave theory (SWT) \cite{IvashkoPhysRevB2017,PengNatPhys2017}. Those have been successful in explaining certain aspects of experiments, yet none of them provides a unified description of both spin and charge excitations within a single microscopic model with fixed parameters. Specifically, DQMC yields well controlled imaginary-time susceptibilities, but suffers from the sign problem and requires analytic continuation of numerical data, which reduces its reliability in regard to dynamics. Moreover, due to lattice-size limitations, DQMC cannot account for long-range Coulomb repulsion that is considered essential for plasmon physics in high-$T_c$ materials  \cite{MarkiewiczPhysRevB2008}. On the other hand, Hubbard-operator large-$N$ limit with long-range interactions included, reproduces measured plasmon spectra  \cite{NagArxiv2020}, but it is intended for the strong-coupling situation ($t$-$J$/$t$-$J$-$V$ models) and seems to overestimate  correlation effects, such as bandwidth renormalization. This has been compensated by adopting bare nearest-neighbor hopping scale $|t| \approx 0.5$-$0.75 \, \mathrm{eV}$  \cite{GrecoCommunPhys2019,GrecoPhysRevB2020,NagArxiv2020}, larger than accepted values $|t| \approx 0.3$-$0.4 \, \mathrm{eV}$. Also, the Hubbard-operator $1/N$ expansion does not treat the collective modes on the same footing and privileges charge- over spin excitations \cite{FoussatsPhysRevB2002,FoussatsPhysRevB2004}. On the other hand, the RPA approach requires adopting unphysically small on-site repulsion $U \sim 1.5 |t|$ \cite{GuariseNatCommun2014,ZhangJPCM2020}. Finally, accurate fits to the paramagnon spectra are obtained by applying SWT to extended Heisenberg models, including both cyclic- and long-range exchange \cite{IvashkoPhysRevB2017,PengNatPhys2017}. Yet, SWT disregards charge excitations, and the underlying large-spin approximation yields magnetic order at high-doping, in disagreement with experiment. In effect, a consistent theoretical picture of spin- and charge dynamics in metallic high-$T_c$ cuprates has not been reached so far.

We fill-in this gap and reconcile quantitatively both paramagnon and plasmon excitations in hole-doped cuprates within a single microscopic model with realistic and doping-independent microscopic parameters. We start from a three-dimensional Hubbard Hamiltonian, with long-range Coulomb repulsion included, and analyze it using recently developed VWF+$1/\mathcal{N}_f$ scheme \cite{FidrysiakPhysRevB2020,FidrysiakArXiv2020} that combines \textbf{V}ariational \textbf{W}ave \textbf{F}unction (VWF) approach with expansion in inverse number of fermionic flavors ($1/\mathcal{N}_f$). This allows us to account for both spin- and charge quantum fluctuations around the correlated ground state \emph{on the same footing}, which is needed for an unbiased analysis. Explicitly, we show that intense and propagating paramagnons persist in the metallic phase along the anti-nodal ($\Gamma$-$X$) Brillouin-zone (BZ) direction in wide doping range, but become rapidly overdamped along the nodal ($\Gamma$-$M$) line. This reflects the experimental trends for multiple copper-oxide families \cite{DeanNatMater2013,IshiiNatCommun2014,LeeNatPhys2014,GuariseNatCommun2014,WakimotoPhysRevB2015,MinolaPhysRevLett2017,IvashkoPhysRevB2017,MeyersPhysRevB2017,ChaixPhysRevB2018,Robarts_arXiV_2019,ZhouNatCommun2013,GretarssonPhysRevLett2016,FumagalliPhysRevB2019,LeTaconNatPhys2011,JiaNatCommun2014,PengPhysRevB2018}. Also, plasmons are shown to exhibit a substantial three-dimensional character. The results agree semi-quantitatively with available RIXS data for $\mathrm{(Bi, Pb)_2 (Sr, La)_2 CuO_{6+\delta}}$. In effect, VWF+$1/\mathcal{N}_f$ emerges as a platform for quantitative interpretation of spectroscopic data for strongly correlated materials, and combines with former equilibrium VWF results \cite{SpalekPhysRevB2017,ZegrodnikPhysRevB2017_2,ZegrodnikPhysRevB2017_3,FidrysiakJPhysCondensMatter2018,ZegrodnikPhysRevB2018,ZegrodnikJPCM2021} into a consistent overall description of high-$T_c$ cuprate superconductors.

\begin{figure}
  \centering
  \includegraphics[width=\linewidth]{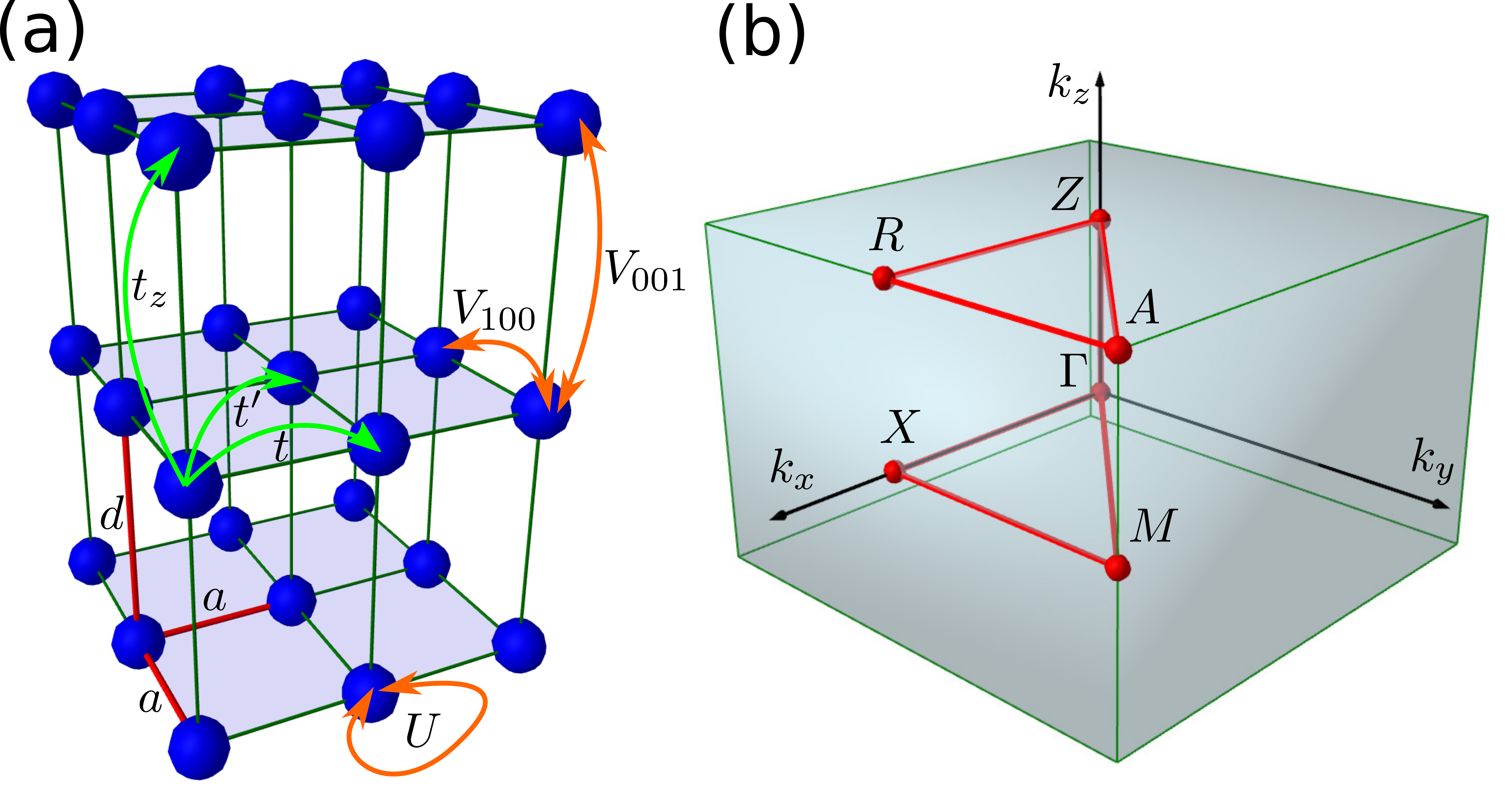}
  \caption{(a) Layered square lattice with in-plane lattice constant $a$ and interlayer spacing $d$. Model parameters are marked inside: green and orange arrows indicate hopping integrals and interactions, respectively. Only two out of infinite number of long-range Coulomb integrals, $V$, are shown. (b) First Brillouin zone with marked $\Gamma$-$X$-$M$-$\Gamma$-$Z$-$R$-$A$-$Z$ contour.} \label{fig:lattice}
\end{figure}

\emph{Model and method}---We employ extended Hubbard model Hamiltonian

\begin{align}
  \label{eq:hamitlonian}
  \hat{\mathcal{H}} = \sum_{ij\sigma} t_{ij} \hat{c}_{i\sigma}^\dagger \hat{c}_{j\sigma} + U \sum_i \hat{n}_{i\uparrow}  \hat{n}_{i\downarrow} + \frac{1}{2}\sum_{i \neq j} V_{ij} \hat{n}_i \hat{n}_j,
\end{align}

\noindent
where $\hat{c}_{i\sigma}^\dagger$ ($\hat{c}_{i\sigma}$) are creation (annihilation) operators on site $i$ for spin $\sigma$, $\hat{n}_{i\sigma} \equiv \hat{c}^\dagger_{i\sigma} \hat{c}_{i\sigma}$, and $\hat{n} \equiv \hat{n}_\downarrow + \hat{n}_\uparrow$. The model is defined on a stacked two-dimensional square lattice ($200 \times 200 \times 16$ sites) with in-plane spacing $a$, and interlayer distance $d$ (cf. Fig.~\ref{fig:lattice}). We adopt standard values of in-plane nearest-neighbor (n.n.) and next-nearest hopping integrals, $t = -0.35\,\mathrm{eV}$ and $t^\prime = 0.25 |t|$, respectively, and small out-of-plane one $t_z = -0.01 |t|$, reflecting substantial interlayer distance in Bi2201. The on-site Coulomb repulsion is set to $U = 6 |t|$, which is backed by recent estimates of effective $U \sim 6$-$9\,|t|$ \cite{NilssonPhysRevB2019}, and has been adopted in a single-layer model study \cite{FidrysiakPhysRevB2020}.  The last term accounts for long-range Coulomb repulsion. At large distances,  $V_{ij}$ may be obtained as a solution of a discretized Laplace equation \cite{BeccaPhysRevB1996}, yielding in $\mathbf{k}$-space

\begin{align}
  \label{eq:long_range_coulomb}
  V_\mathbf{k} = \frac{V_c}{\gamma \Phi(k_x, k_y) + 1 - \cos\left(k_z d\right)},
\end{align}

\begin{figure}
  \centering
  \includegraphics[width=\linewidth]{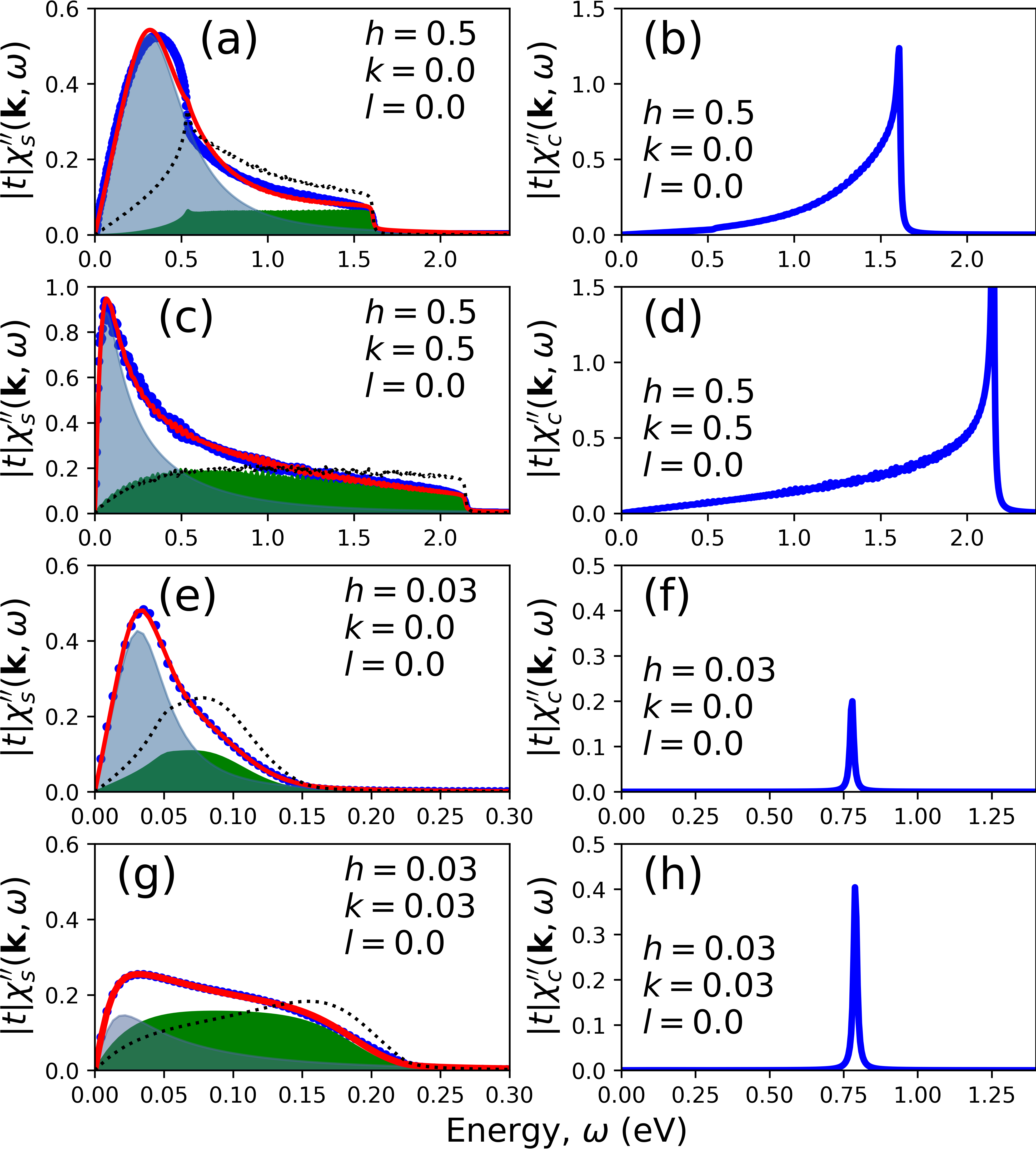}
  \caption{Representative $\mathrm{SGA}_f$+$1/\mathcal{N}_f$ results for $\delta=0.16$. Blue points and lines are the calculated imaginary parts of spin [(a), (c), (e), (g)] and charge [(b), (d), (f), (h)] dynamical susceptibilities for wave vectors detailed inside the panels. On the left, the green-shaded areas represent incoherent contribution, whereas blue-shaded area is the harmonic oscillator peak [cf. Eq.~\eqref{eq:spin_susceptibility_fit}]. The red line is the sum of the two, and black dotted lines represent Lindhard response. Note emergence of the coherent plasmon peaks close to the BZ center.} \label{fig:fitting_procedure}
\end{figure}

\noindent
where $\Phi(k_x, k_y) \equiv 2 - \cos(k_xa) - \cos(k_ya)$, $V_c = e^2 d / (2 a^2\epsilon_\perp \epsilon_0)$, $\gamma = \epsilon_{||} d^2 / (\epsilon_\perp a^2)$, and $\epsilon_{||}$ ($\epsilon_\perp$) are in-plane (out of-plane) dielectric constants. We select $\gamma = 10$ and $V_c = 46 |t|$, which assumes a dominant lattice-anisotropy effect on $\gamma$ (see \cite{NagArxiv2020}) and yields $\epsilon_\perp \approx 4.66$, comparable with the high-energy values $\epsilon_\perp \approx 4$-$4.5$ reported for Bi2201 \cite{vanHeumenN2009}. Also, the resulting n.n. repulsion $V/|t| \approx 2.03 $ is consistent with \emph{ab initio} \cite{HirayamaPhysRevB2018} estimates for related materials ($2.36$ for $\mathrm{HgBa_2CuO_4}$ and $2.30$ for $\mathrm{La_2 CuO_4}$). Note that the plasmon gap can be estimated as $\Delta_p^2 = 2 \hbar^2 V_c n_c / (m^{*} a^2 \gamma)$, with $m^{*}$ and $n_c$ being the correlation-renormalized carrier mass and concentration, respectively. The scale of $\Delta_p$ is thus sensitive to $V_c/\gamma = 4.6 |t|$. Hereafter we set the temperature to $k_BT = 0.4 |t|$ to stay clear of broken-symmetry \cite{IgoshevJPCM2015} states.

The model~\eqref{eq:hamitlonian} is solved using VWF+$1/\mathcal{N}_f$ scheme which has been elaborated extensively in a methodological paper \cite{FidrysiakArXiv2020}, regarded here as a Supplemental Material. In brief, the method is based on the energy functional $E_\mathrm{var} \equiv \langle\Psi_\mathrm{var}|\hat{\mathcal{H}}|\Psi_\mathrm{var}\rangle/\langle\Psi_\mathrm{var}|\Psi_\mathrm{var}\rangle$, defined in terms of the variational state $|\Psi_\mathrm{var}\rangle \equiv \hat{P}_\mathrm{var}(\boldsymbol{\lambda}) |\Psi_0\rangle$, where $|\Psi_0\rangle$ is an uncorrelated wave function. The operator $\hat{P}_\mathrm{var}(\boldsymbol{\lambda})$ adjusts weights of many-body configurations in $|\Psi_\mathrm{var}\rangle$ and depends on a vector composed of variational parameters, $\boldsymbol{\lambda}$, subjected to additional constraints \cite{FidrysiakArXiv2020}. By application of linked-cluster expansion in real space, $E_\mathrm{var} = E_\mathrm{var}(\mathbf{P}, \boldsymbol{\lambda})$ becomes a functional of ``lines'', $P_{i\sigma j\sigma^\prime} \equiv \langle\hat{c}_{i\sigma}^\dagger \hat{c}_{j\sigma^\prime}\rangle$, and $\boldsymbol{\lambda}$. We use \textbf{S}tatistically-consistent \textbf{G}utzwiller \textbf{A}pproximation (SGA)\cite{JedrakPhsRevB2011} to truncate diagrammatic series for $E_\mathrm{var}$, which results in $\mathrm{SGA}_f$+$1/\mathcal{N}_f$ variant of VWF+$1/\mathcal{N}_f$. We also approximate long-range part of Coulomb energy as $\langle{\hat{V}}\rangle \approx \frac{1}{2}\sum_{i \neq j} V_{ij} \langle{n_i}\rangle \langle{n_j}\rangle$, effectively disregarding non-local lines which is justified at large distances. As a second step, $\mathbf{P} \rightarrow \mathbf{P}(\tau)$ and $\boldsymbol{\lambda} \rightarrow \boldsymbol{\lambda}(\tau)$ are promoted to (imaginary-time) dynamical fields. Finally, the Euclidean action for $\mathbf{P}(\tau)$, $\boldsymbol{\lambda}(\tau)$, and other auxiliary fields, is constructed and used to generate dynamical spin- and charge- collective susceptibilities, $\chi_s(\mathbf{k}, i \omega_n)$ and $\chi_c(\mathbf{k}, i \omega_n)$, respectively. Analytic continuation is carried out as $i \omega_n \rightarrow \omega + i 0.02 |t|$.

To make a comparison with experiment, it is necessary to extract paramagnon energies and their damping rates from the calculated spectra. This is done by damped harmonic oscillator modeling \cite{LamsalPhysRevB2016} of the imaginary part of the dynamical spin susceptibility

\begin{align}
  \label{eq:spin_susceptibility_fit}
  \chi^{\prime\prime}_s(\mathbf{k}, \omega) = \frac{2 A(\mathbf{k}) \gamma(\mathbf{k}) \omega}{ \left[\omega^2 - \omega^2_0(\mathbf{k})\right]^2 + 4 \gamma^2(\mathbf{k}) \omega^2} + \chi^{\prime\prime}_{s, \mathrm{in}}(\mathbf{k}, \omega),
\end{align}

\noindent
where $A(\mathbf{k})$, $\omega_0(\mathbf{k})$, and $\gamma(\mathbf{k})$ denote the amplitude, bare energy, and damping rate, respectively. Crucially, $\omega_0(\mathbf{k})$ \emph{does not} represent the physical paramagnon energy, and it remains non-zero even if magnetic excitations are \textit{overdamped}. The relevant parameter is thus the real part of the quasiparticle pole, $\omega_p(\mathbf{k}) = \sqrt{\omega^2_0(\mathbf{k}) - \gamma^2(\mathbf{k})}$ if $\omega_0(\mathbf{k}) > \gamma(\mathbf{k})$, and zero otherwise. The last term represents the incoherent part, $\chi^{\prime\prime}_{s, \mathrm{in}}(\mathbf{k}, \omega)$, providing background to the oscillator peak. We model $\chi^{\prime\prime}_{s, \mathrm{in}}(\mathbf{k}, \omega)$ as the Lindhard susceptibility (defined as the loop integral, evaluated with Landau quasiparticle Green's functions), multiplied by a linear function of $\omega$ to allow for spectral-weight redistribution between the coherent- and incoherent parts. Thus, $\chi^{\prime\prime}_{s, \mathrm{in}}(\mathbf{k}, \omega) \equiv [B(\mathbf{k}) + \omega C(\mathbf{k})] \cdot \chi^{\prime\prime}_{0s}(\mathbf{k}, \omega)$, where $B(\mathbf{k}) \geq 0$ and $C(\mathbf{k})$ are free $\mathbf{k}$-dependent parameters. This form of $\chi^{\prime\prime}_{s, \mathrm{in}}(\mathbf{k}, \omega)$ reflects the fermiology of the underlying correlated electronic system.

\emph{Results}---Representative least-squares fits of imaginary parts of the $\mathrm{SGA}_f$+$1/\mathcal{N}_f$ susceptibilities over the energy range approximately encompassing non-zero values of $\chi_s^{\prime\prime}(\mathbf{k}, \omega)$, are displayed in left panels of Fig.~\ref{fig:fitting_procedure}. Blue circles in (a), (c), (e) and (g) represent calculated $\chi_s^{\prime\prime}(\mathbf{k}, \omega)$ for $\mathbf{k} = (0.5, 0, 0)$, $(0.5, 0.5, 0)$, $(0.03, 0, 0)$, and $(0.03, 0.03, 0)$, respectively. The green and blue regions are the incoherent and harmonic parts, respectively. The red line marks sum of the two, reproducing faithfully the $\mathrm{SGA}_f$+$1/\mathcal{N}_f$ result. For completeness, by black dotted lines we depict the Lindhard susceptibility, character of which varies across the BZ. A substantial directional anisotropy of spin dynamics is apparent, with coherent oscillator peaks appearing only along the anti-nodal line. In the right panels, the corresponding charge response, $\chi_c^{\prime\prime}(\mathbf{k}, \omega)$, is plotted. A clear distinction between the incoherent part and plasmon peak should be noted for the experimentally relevant regime of small in-plane momentum transfers, hence we identify the plasmon energy with the peak position in $\chi^{\prime\prime}_c(\mathbf{k}, \omega)$.

\begin{figure*}
\centering
  \includegraphics[width=0.9\linewidth]{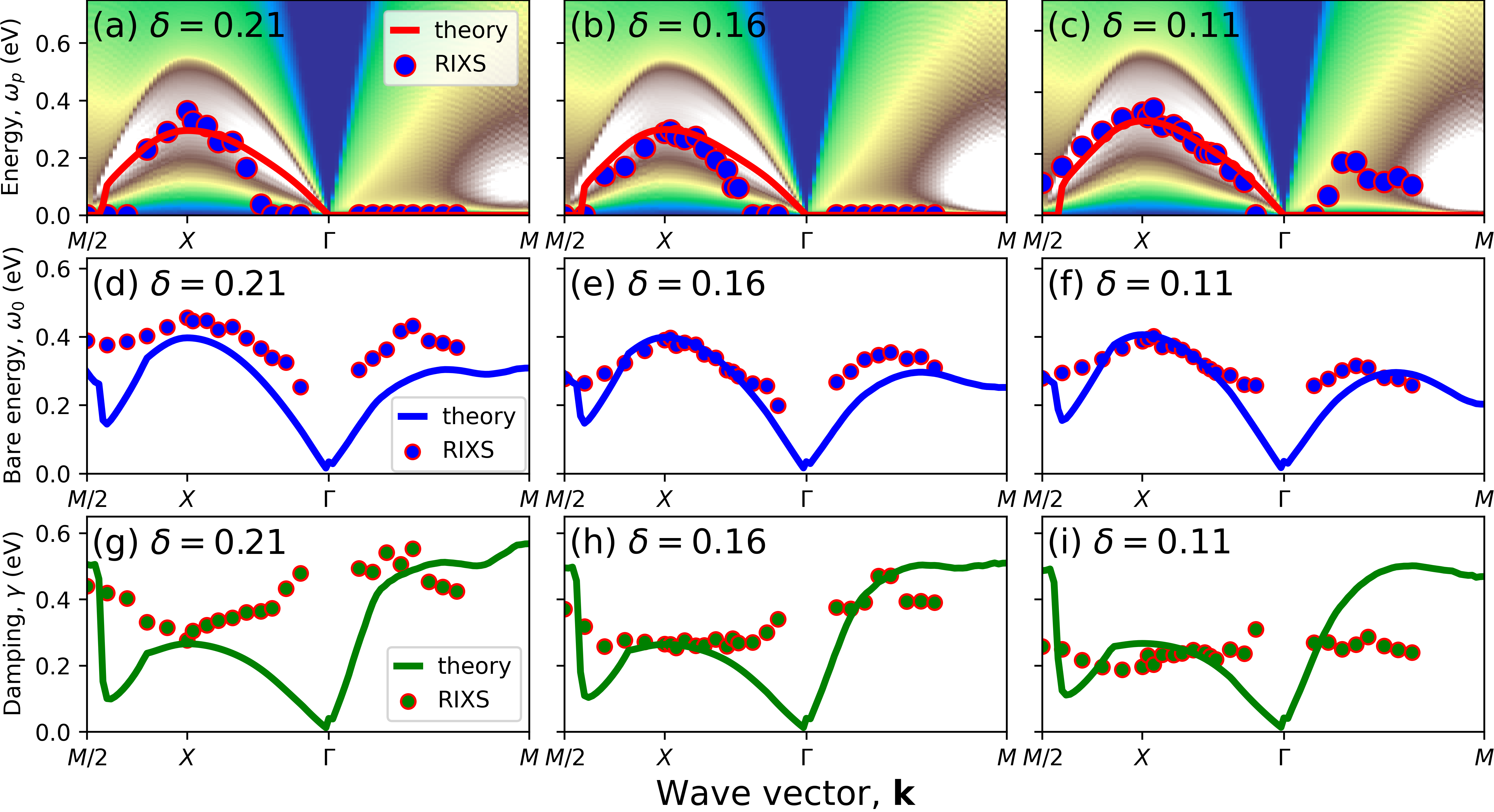}
  \caption{Paramagnon energies for $\mathrm{(Bi, Pb)_2 (Sr, La)_2 CuO_{6+\delta}}$ along the $M/2$-$X$-$\Gamma$-$M$ Brillouin-zone contour. To panels [(a)-(c)] show the propagation energy, $\omega_p$, as obtained from $\mathrm{SGA}_f$+$1/\mathcal{N}_f$ approach (red solid lines) and experiment\cite{PengPhysRevB2018} (circles) at three hole-doping levels, $\delta=0.21$, $\delta=0.16$, and $\delta=0.11$. The color maps represent calculated imaginary part of dynamical spin susceptibility, ranging from blue (low intensity) to white (high intensity). Panels (d)-(f) and (g)-(i) show the bare paramagnon energies and damping rate ($\omega_0$ and $\gamma$, respectively). Lines and circles are $\mathrm{SGA}_f$+$1/\mathcal{N}_f$ results and RIXS data \cite{PengPhysRevB2018}, respectively. }
  \label{fig:paramagnons}
\end{figure*}

We now proceed to a unified quantitative analysis of both paramagnon and plasmon dynamics in $\mathrm{(Bi, Pb)_2 (Sr, La)_2 CuO_{6+\delta}}$. In Fig.~\ref{fig:paramagnons} we compare the calculated $\mathrm{SGA}_f$+$1/\mathcal{N}_f$ paramagnon characteristics with RIXS data for hole-doping levels $\delta=0.21$ [(a), (d), (g)], $\delta=0.16$ [(b), (e), (h)], and $\delta=0.11$ [(c), (f), (i)]. Top panels [(a)-(c)] show the $\mathrm{SGA}_f$+$1/\mathcal{N}_f$ (red solid lines) and RIXS \cite{PengPhysRevB2018} (solid circles) paramagnon propagation energies, $\omega_p(\mathbf{k})$. Color maps represent imaginary part of the dynamical spin susceptibility, with blue and white colors mapping to low- and high-intensity regions, respectively. The agreement between theory and experiment is semi-quantitative for all doping levels, with the exception of the $\Gamma$-$M$ direction for $\delta = 0.11$. In the latter case, $\mathrm{SGA}_f$+$1/\mathcal{N}_f$ yields overdamped magnetic dynamics ($\omega_p(\mathbf{k}) = 0$), whereas RIXS data corresponds to substantially damped, but still resonant response. A significant anisotropy between the nodal ($\Gamma$-$M$) and anti-nodal ($\Gamma$-$X$) directions is consistently observed both in $\mathrm{SGA}_f$+$1/\mathcal{N}_f$ and experimental data. Namely, the anti-nodal paramagnons persist across the entire doping range, but become rapidly overdamped with increasing doping along the nodal line. We note that a comparable agreement with RIXS paramagnon spectra has been recently achieved within a single-layer model \cite{FidrysiakPhysRevB2020}. This points towards a predominately two-dimensional character of spin excitations, which is also supported by investigation of the static response, detailed below.

In Fig.~\ref{fig:paramagnons}(d)-(f) and \ref{fig:paramagnons}(g)-(i), we carry out an analysis of the underlying bare paramagnon energies and damping, $\omega_0(\mathbf{k})$ and $\gamma(\mathbf{k})$, respectively. Solid lines mark the parameters extracted from $\mathrm{SGA}_f$+$1/\mathcal{N}_f$ spin susceptibilities, with the use of model~\eqref{eq:spin_susceptibility_fit}, whereas solid circles are RIXS data of Ref.~\cite{PengPhysRevB2018}, processed in an analogous manner. The overall agreement of both quantities with experiment is semi-quantitative across the phase diagram, with the exception of the $\Gamma$-$M$ line in the underdoped case, where $\mathrm{SGA}_f$+$1/\mathcal{N}_f$ yields larger damping rates, and close to the $\Gamma$ point for the overdoped situation.

\begin{figure}
  \centering
  \includegraphics[width=\linewidth]{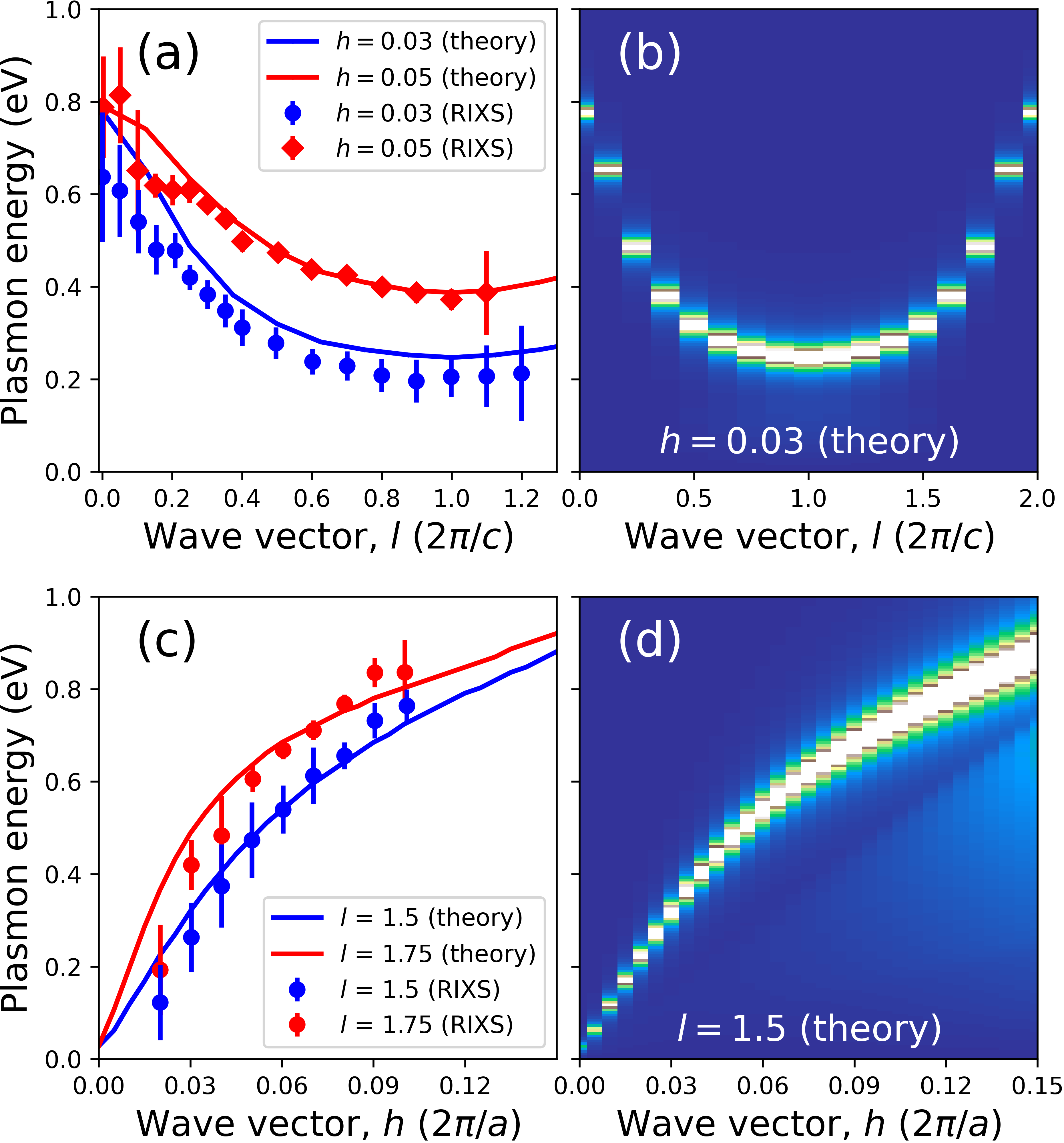}
  \caption{Plasmon dispersion in $\mathrm{Bi_2Sr_{1.6}La_{0.4}CuO_{6+\delta}}$ at doping $\delta=0.16$. Lines and symbols are $\mathrm{SGA}_f$+$1/\mathcal{N}_f$ results and the RIXS data of Ref.~\cite{NagArxiv2020}, respectively. Panel (a) shows the data as a function the out-of-plane momentum transfer, $l$, for fixed $h=0.03$ (blue color) and $h=0.05$ (red color). In panel (c), two in-plane cuts for $l=1.5$ (blue) and $l=1.75$ (red) are displayed. In (b) and (d), the corresponding raw $\mathrm{SGA}_f$+$1/\mathcal{N}_f$ dynamical charge susceptibilities are shown. Wave vectors are expressed as $\mathbf{k} = (h \frac{2\pi}{a}, 0, l \frac{2 \pi}{c})$, with $c = 2d$.}
  \label{fig:plasmons}
\end{figure}
\begin{figure}[h]
  \centering
  \includegraphics[width=\linewidth]{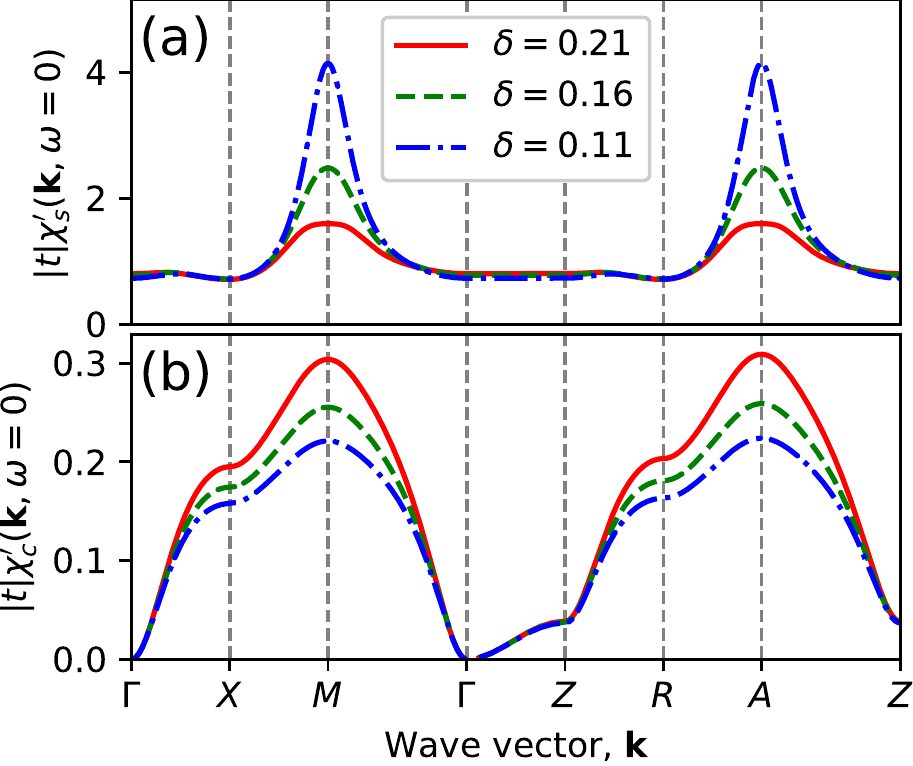}
  \caption{Calculated static spin [panel (a)] and charge [panel(b)] susceptibilities for hole doping $\delta = 0.21$ (red solid lines), $\delta = 0.16$ (green dashed line), and $\delta = 0.11$ (blue dash-dotted line). The universal feature at point $M$ [in the case (a)] and at $\Gamma$ [in the case (b)] should be noted.}
  \label{fig:stability}
\end{figure}

We turn next to the discussion of charge excitations for \emph{the same model parameters} as those used to generate Fig.~\ref{fig:paramagnons}. The wave-vector transfers are hereafter represented as $\mathbf{k} = \left(h \frac{2\pi}{a}, 0, l \frac{2\pi}{c}\right)$, with $c$ taken as $2d$ to account for two primitive cells in a crystallographic cell \cite{KovalevaPhysRevB2004}. In Fig.~\ref{fig:plasmons}(a), the calculated plasmon energies as a function of $l$ are displayed for $h = 0.03$ (blue line) and $h = 0.05$ (red line), and compared with the corresponding RIXS data \cite{NagArxiv2020} for $\mathrm{Bi_2Sr_{1.6}La_{0.4}CuO_{6+\delta}}$. For reference, in panel (b) we show raw imaginary part of the $\mathrm{SGA}_{\lambda_d}$+$1/\mathcal{N}_f$ charge susceptibility for $h = 0.03$, used to obtain theoretical dispersion curve [blue line in (a)]. Panel (c) exhibits in-plane plasmon dispersion relations for two fixed values of the out-of-plane wave-vector transfer, $l=1.5$ and $l=1.75$, as a function of $h$. In panel (d), we display unprocessed $\chi_c^{\prime\prime}$ for $l=1.5$. The agreement between theory and experiment is quantitative along all BZ contours. As is seen in Fig.~\ref{fig:plasmons}(a), plasmon modes disperse strongly along the out-of-plane direction, with the minimum energy for $l = 1$, corresponding to anti-phase charge fluctuations on neighboring $\mathrm{CuO_2}$ planes.

For the sake of completeness, we also examine stability of the paramagnetic metallic state against fluctuations. In Fig.~\ref{fig:stability}, static spin [panel (a)] and charge [panel (b)] susceptibilities are displayed along the $\Gamma$-$X$-$M$-$\Gamma$-$Z$-$R$-$A$-$Z$ contour (cf. Fig.~\ref{fig:lattice}) for the three doping levels, $\delta=0.21$, $0.16$, and $0.11$. The susceptibilities remain finite along the high-symmetry directions, with an increasing tendency towards antiferromagnetic instability as half-filling is approached [enhanced values of $\chi_s^\prime(\mathbf{k}, \omega=0)$ at the $M$ point; cf. panel (a)]. On the other hand, the charge response [panel (b)] depends only weakly on doping, indicating that the system stays clear of charge-density-wave (CDW) order in the considered temperature range. As is seen in Fig.~\ref{fig:stability}, spin fluctuations are two-dimensional with barely distinguishable $\Gamma$-$X$-$M$-$\Gamma$ and $Z$-$R$-$A$-$Z$ profiles, whereas charge response exhibits qualitatively distinct behavior around $\Gamma$ and $Z$ points. Our results support the physical picture of at most moderate screening of the non-local Coulomb interaction, so that the plasmon excitations are influenced by its algebraic tail. On the other hand, the paramagnons are weakly affected by the non-local terms. The three-dimensional extension of the Hubbard model with inclusion of the long-range interactions is thus required primarily to quantitatively describe charge excitations. The impact of those terms on equilibrium properties has been discussed elsewhere \cite{ZegrodnikJPCM2021}.

\emph{Outlook}---We have carried out a quantitative analysis of collective spin- and charge excitations in a microscopic model of high-$T_c$ copper-oxides. Those modes are present in wide temperature and doping range and, in particular, in the regime where no long-range spin-density-wave or CDW order occur. The principal difficulty in describing them is due to the strongly-correlated character of the underlying electronic states. This circumstance necessitates a generalization of Moriya-Hertz-Millis-type approach to incorporate fluctuations into a nonstandard reference state and going systematically beyond the renormalized mean-field theory (RMFT) \cite{FidrysiakArXiv2020}. The dynamical effects are included by $1/\mathcal{N}_f$ expansion around the variationally-determined saddle-point solution, reproducing the experimental data semi-quantitatively within a single scheme and for once fixed microscopic parameters [cf. Figs.~\ref{fig:paramagnons}(a-c) and ~\ref{fig:plasmons}(a-b)]. In conjunction with the former comprehensive VWF analysis of the static- and single-particle properties of high-$T_c$ cuprates \cite{SpalekPhysRevB2017,ZegrodnikPhysRevB2017_2,ZegrodnikPhysRevB2017_3,FidrysiakJPhysCondensMatter2018,ZegrodnikPhysRevB2018,ZegrodnikJPCM2021}, encompassing SC/CDW phases, Fermi-velocity/wave-vector, quasiparticle masses, and kinetic energy gain at SC transition, we arrive here at a consistent semi-quantitative description of both static- and collective dynamic properties of hole-doped high-$T_c$ materials. Those aspects should be studied further within a more realistic three-band model of high-$T_c$ SC, either in the Hubbard or $t$-$J$-$U$-$V$ form \cite{SpalekPhysRevB2017}.

The untouched here questions comprise pseudogap formation and temperature-dependence of electrical resistivity, when the quantum fluctuations are tackled explicitly along the lines presented here. This requires supplementing the present approach with calculations of single-particle self-energy and subleading fluctuation free-energy corrections, all in a fully self-consistent manner. Such a task poses a substantial challenge. Finally, within the \emph{strong-correlation picture}, both the real-space pairing and AF correlations in the cuprates share the same source: kinetic exchange interaction $\propto \hat{\mathbf{S}}_i \hat{\mathbf{S}}_j - \frac{1}{4} \hat{n}_i\hat{n}_j$, that may be equivalently expressed in terms of singlet pairing operators $\hat{b}^\dagger_{ij} \equiv \frac{1}{\sqrt{2}} (\hat{c}^\dagger_{i\uparrow} \hat{c}^\dagger_{j\downarrow} - \hat{c}^\dagger_{i\downarrow} \hat{c}^\dagger_{j\uparrow})$ \cite{SpalekPhysRevB1988}. The considered here paramagnetic ground state is also spin singlet, and the elementary paramagnon excitations are associated with singlet-triplet ($S=0$ to $S=1$) transitions. Their robustness in hole-doped cuprates supports thus indirectly also the exchange-driven real-space pairing viewpoint, calling for an extension of the VWF+$1/\mathcal{N}_f$ approach to incorporate the SC state. This requires accounting for the SC gap fluctuations through the anomalous lines, $S_{i\sigma j\sigma^\prime} = \langle \hat{c}_{i\sigma} \hat{c}_{j\sigma^\prime}\rangle$ \cite{SpalekPhysRevB2017,FidrysiakJPhysCondensMatter2018,AbramJPhysCondensMatter2017}, introducing additional complexity to the problem, and should be treated separately.

To recapitulate, the VWF solution going systematically beyond RMFT (DE-GWF scheme \cite{SpalekPhysRevB2017,ZegrodnikJPCM2021,ZegrodnikPhysRevB2017_2,ZegrodnikPhysRevB2017_3,FidrysiakJPhysCondensMatter2018,ZegrodnikPhysRevB2018}), combined with the present VWF+$1/\mathcal{N}_f$ approach, supports in a quantitative manner a mutual relationship between strong-electronic correlations and collective dynamics in high-$T_c$ cuprates.

\emph{Acknowledgments}---This work  was  supported  by  Grant  OPUS  No.  UMO-2018/29/B/ST3/02646 from Narodowe Centrum Nauki and by a grant from the SciMat Priority Research Area under the Strategic Programme Excellence Initiative at the Jagiellonian University.
 

%


\begin{thebibliography}{55}%
\makeatletter
\providecommand \@ifxundefined [1]{%
 \@ifx{#1\undefined}
}%
\providecommand \@ifnum [1]{%
 \ifnum #1\expandafter \@firstoftwo
 \else \expandafter \@secondoftwo
 \fi
}%
\providecommand \@ifx [1]{%
 \ifx #1\expandafter \@firstoftwo
 \else \expandafter \@secondoftwo
 \fi
}%
\providecommand \natexlab [1]{#1}%
\providecommand \enquote  [1]{``#1''}%
\providecommand \bibnamefont  [1]{#1}%
\providecommand \bibfnamefont [1]{#1}%
\providecommand \citenamefont [1]{#1}%
\providecommand \href@noop [0]{\@secondoftwo}%
\providecommand \href [0]{\begingroup \@sanitize@url \@href}%
\providecommand \@href[1]{\@@startlink{#1}\@@href}%
\providecommand \@@href[1]{\endgroup#1\@@endlink}%
\providecommand \@sanitize@url [0]{\catcode `\\12\catcode `\$12\catcode
  `\&12\catcode `\#12\catcode `\^12\catcode `\_12\catcode `\%12\relax}%
\providecommand \@@startlink[1]{}%
\providecommand \@@endlink[0]{}%
\providecommand \url  [0]{\begingroup\@sanitize@url \@url }%
\providecommand \@url [1]{\endgroup\@href {#1}{\urlprefix }}%
\providecommand \urlprefix  [0]{URL }%
\providecommand \Eprint [0]{\href }%
\providecommand \doibase [0]{http://dx.doi.org/}%
\providecommand \selectlanguage [0]{\@gobble}%
\providecommand \bibinfo  [0]{\@secondoftwo}%
\providecommand \bibfield  [0]{\@secondoftwo}%
\providecommand \translation [1]{[#1]}%
\providecommand \BibitemOpen [0]{}%
\providecommand \bibitemStop [0]{}%
\providecommand \bibitemNoStop [0]{.\EOS\space}%
\providecommand \EOS [0]{\spacefactor3000\relax}%
\providecommand \BibitemShut  [1]{\csname bibitem#1\endcsname}%
\let\auto@bib@innerbib\@empty
\bibitem [{\citenamefont {Keimer}\ \emph {et~al.}(2015)\citenamefont {Keimer},
  \citenamefont {Kivelson}, \citenamefont {Norman}, \citenamefont {Uchida},\
  and\ \citenamefont {Zaanen}}]{KeimerNature2015}%
  \BibitemOpen
  \bibfield  {author} {\bibinfo {author} {\bibfnamefont {B.}~\bibnamefont
  {Keimer}}, \bibinfo {author} {\bibfnamefont {S.~A.}\ \bibnamefont
  {Kivelson}}, \bibinfo {author} {\bibfnamefont {M.~R.}\ \bibnamefont
  {Norman}}, \bibinfo {author} {\bibfnamefont {S.}~\bibnamefont {Uchida}}, \
  and\ \bibinfo {author} {\bibfnamefont {J.}~\bibnamefont {Zaanen}},\
  }\bibfield  {title} {\enquote {\bibinfo {title} {{From quantum matter to
  high-temperature superconductivity in copper oxides}},}\ }\href {\doibase
  10.1038/nature14165} {\bibfield  {journal} {\bibinfo  {journal} {Nature}\
  }\textbf {\bibinfo {volume} {518}},\ \bibinfo {pages} {179} (\bibinfo {year}
  {2015})}\BibitemShut {NoStop}%
\bibitem [{\citenamefont {Coldea}\ \emph {et~al.}(2001)\citenamefont {Coldea},
  \citenamefont {Hayden}, \citenamefont {Aeppli}, \citenamefont {Perring},
  \citenamefont {Frost}, \citenamefont {Mason}, \citenamefont {Cheong},\ and\
  \citenamefont {Fisk}}]{ColdeaPhysRevLett2001}%
  \BibitemOpen
  \bibfield  {author} {\bibinfo {author} {\bibfnamefont {R.}~\bibnamefont
  {Coldea}}, \bibinfo {author} {\bibfnamefont {S.~M.}\ \bibnamefont {Hayden}},
  \bibinfo {author} {\bibfnamefont {G.}~\bibnamefont {Aeppli}}, \bibinfo
  {author} {\bibfnamefont {T.~G.}\ \bibnamefont {Perring}}, \bibinfo {author}
  {\bibfnamefont {C.~D.}\ \bibnamefont {Frost}}, \bibinfo {author}
  {\bibfnamefont {T.~E.}\ \bibnamefont {Mason}}, \bibinfo {author}
  {\bibfnamefont {S.-W.}\ \bibnamefont {Cheong}}, \ and\ \bibinfo {author}
  {\bibfnamefont {Z.}~\bibnamefont {Fisk}},\ }\bibfield  {title} {\enquote
  {\bibinfo {title} {{Spin Waves and Electronic Interactions in
  ${\mathrm{La}}_{2}{\mathrm{CuO}}_{4}$}},}\ }\href {\doibase
  10.1103/PhysRevLett.86.5377} {\bibfield  {journal} {\bibinfo  {journal}
  {Phys. Rev. Lett.}\ }\textbf {\bibinfo {volume} {86}},\ \bibinfo {pages}
  {5377} (\bibinfo {year} {2001})}\BibitemShut {NoStop}%
\bibitem [{\citenamefont {Peng}\ \emph {et~al.}(2017)\citenamefont {Peng},
  \citenamefont {Dellea}, \citenamefont {Minola}, \citenamefont {Conni},
  \citenamefont {Amorese}, \citenamefont {Di~Castro}, \citenamefont {De~Luca},
  \citenamefont {Kummer}, \citenamefont {Salluzzo}, \citenamefont {Sun},
  \citenamefont {Zhou}, \citenamefont {Balestrino}, \citenamefont {Le~Tacon},
  \citenamefont {Keimer}, \citenamefont {Braicovich}, \citenamefont {Brookes},\
  and\ \citenamefont {Ghiringhelli}}]{PengNatPhys2017}%
  \BibitemOpen
  \bibfield  {author} {\bibinfo {author} {\bibfnamefont {Y.~Y.}\ \bibnamefont
  {Peng}}, \bibinfo {author} {\bibfnamefont {G.}~\bibnamefont {Dellea}},
  \bibinfo {author} {\bibfnamefont {M.}~\bibnamefont {Minola}}, \bibinfo
  {author} {\bibfnamefont {M.}~\bibnamefont {Conni}}, \bibinfo {author}
  {\bibfnamefont {A.}~\bibnamefont {Amorese}}, \bibinfo {author} {\bibfnamefont
  {D.}~\bibnamefont {Di~Castro}}, \bibinfo {author} {\bibfnamefont {G.~M.}\
  \bibnamefont {De~Luca}}, \bibinfo {author} {\bibfnamefont {K.}~\bibnamefont
  {Kummer}}, \bibinfo {author} {\bibfnamefont {M.}~\bibnamefont {Salluzzo}},
  \bibinfo {author} {\bibfnamefont {X.}~\bibnamefont {Sun}}, \bibinfo {author}
  {\bibfnamefont {X.~J.}\ \bibnamefont {Zhou}}, \bibinfo {author}
  {\bibfnamefont {G.}~\bibnamefont {Balestrino}}, \bibinfo {author}
  {\bibfnamefont {M.}~\bibnamefont {Le~Tacon}}, \bibinfo {author}
  {\bibfnamefont {B.}~\bibnamefont {Keimer}}, \bibinfo {author} {\bibfnamefont
  {L.}~\bibnamefont {Braicovich}}, \bibinfo {author} {\bibfnamefont {N.~B.}\
  \bibnamefont {Brookes}}, \ and\ \bibinfo {author} {\bibfnamefont
  {G.}~\bibnamefont {Ghiringhelli}},\ }\bibfield  {title} {\enquote {\bibinfo
  {title} {{Influence of apical oxygen on the extent of in-plane exchange
  interaction in cuprate superconductors}},}\ }\href
  {http://dx.doi.org/10.1038/nphys4248} {\bibfield  {journal} {\bibinfo
  {journal} {Nat. Phys.}\ }\textbf {\bibinfo {volume} {13}},\ \bibinfo {pages}
  {1201} (\bibinfo {year} {2017})}\BibitemShut {NoStop}%
\bibitem [{\citenamefont {Dean}\ \emph {et~al.}(2013)\citenamefont {Dean},
  \citenamefont {Dellea}, \citenamefont {Springell}, \citenamefont
  {Yakhou-Harris}, \citenamefont {Kummer}, \citenamefont {Brookes},
  \citenamefont {Liu}, \citenamefont {Sun}, \citenamefont {Strle},
  \citenamefont {Schmitt}, \citenamefont {Braicovich}, \citenamefont
  {Ghiringhelli}, \citenamefont {Bo\v{z}ovi\'c},\ and\ \citenamefont
  {Hill}}]{DeanNatMater2013}%
  \BibitemOpen
  \bibfield  {author} {\bibinfo {author} {\bibfnamefont {M.~P.~M.}\
  \bibnamefont {Dean}}, \bibinfo {author} {\bibfnamefont {G.}~\bibnamefont
  {Dellea}}, \bibinfo {author} {\bibfnamefont {R.~S.}\ \bibnamefont
  {Springell}}, \bibinfo {author} {\bibfnamefont {F.}~\bibnamefont
  {Yakhou-Harris}}, \bibinfo {author} {\bibfnamefont {K.}~\bibnamefont
  {Kummer}}, \bibinfo {author} {\bibfnamefont {N.~B.}\ \bibnamefont {Brookes}},
  \bibinfo {author} {\bibfnamefont {X.}~\bibnamefont {Liu}}, \bibinfo {author}
  {\bibfnamefont {Y-J.}\ \bibnamefont {Sun}}, \bibinfo {author} {\bibfnamefont
  {J.}~\bibnamefont {Strle}}, \bibinfo {author} {\bibfnamefont
  {T.}~\bibnamefont {Schmitt}}, \bibinfo {author} {\bibfnamefont
  {L.}~\bibnamefont {Braicovich}}, \bibinfo {author} {\bibfnamefont
  {G.}~\bibnamefont {Ghiringhelli}}, \bibinfo {author} {\bibfnamefont
  {I.}~\bibnamefont {Bo\v{z}ovi\'c}}, \ and\ \bibinfo {author} {\bibfnamefont
  {J.~P.}\ \bibnamefont {Hill}},\ }\bibfield  {title} {\enquote {\bibinfo
  {title} {{Persistence of magnetic excitations in
  $\mathrm{La_{2-\mathit{x}}Sr_\mathit{x}CuO_4}$ from the undoped insulator to
  the heavily overdoped non-superconducting metal}},}\ }\href
  {http://dx.doi.org/10.1038/nmat3723} {\bibfield  {journal} {\bibinfo
  {journal} {Nat. Mater.}\ }\textbf {\bibinfo {volume} {12}},\ \bibinfo {pages}
  {1019} (\bibinfo {year} {2013})}\BibitemShut {NoStop}%
\bibitem [{\citenamefont {Ishii}\ \emph {et~al.}(2014)\citenamefont {Ishii},
  \citenamefont {Fujita}, \citenamefont {Sasaki}, \citenamefont {Minola},
  \citenamefont {Dellea}, \citenamefont {Mazzoli}, \citenamefont {Kummer},
  \citenamefont {Ghiringhelli}, \citenamefont {Braicovich}, \citenamefont
  {Tohyama}, \citenamefont {Tsutsumi}, \citenamefont {Sato}, \citenamefont
  {Kajimoto}, \citenamefont {Ikeuchi}, \citenamefont {Yamada}, \citenamefont
  {Yoshida}, \citenamefont {Kurooka},\ and\ \citenamefont
  {Mizuki}}]{IshiiNatCommun2014}%
  \BibitemOpen
  \bibfield  {author} {\bibinfo {author} {\bibfnamefont {K.}~\bibnamefont
  {Ishii}}, \bibinfo {author} {\bibfnamefont {M.}~\bibnamefont {Fujita}},
  \bibinfo {author} {\bibfnamefont {T.}~\bibnamefont {Sasaki}}, \bibinfo
  {author} {\bibfnamefont {M.}~\bibnamefont {Minola}}, \bibinfo {author}
  {\bibfnamefont {G.}~\bibnamefont {Dellea}}, \bibinfo {author} {\bibfnamefont
  {C.}~\bibnamefont {Mazzoli}}, \bibinfo {author} {\bibfnamefont
  {K.}~\bibnamefont {Kummer}}, \bibinfo {author} {\bibfnamefont
  {G.}~\bibnamefont {Ghiringhelli}}, \bibinfo {author} {\bibfnamefont
  {L.}~\bibnamefont {Braicovich}}, \bibinfo {author} {\bibfnamefont
  {T.}~\bibnamefont {Tohyama}}, \bibinfo {author} {\bibfnamefont
  {K.}~\bibnamefont {Tsutsumi}}, \bibinfo {author} {\bibfnamefont
  {K.}~\bibnamefont {Sato}}, \bibinfo {author} {\bibfnamefont {R.}~\bibnamefont
  {Kajimoto}}, \bibinfo {author} {\bibfnamefont {K.}~\bibnamefont {Ikeuchi}},
  \bibinfo {author} {\bibfnamefont {K.}~\bibnamefont {Yamada}}, \bibinfo
  {author} {\bibfnamefont {M.}~\bibnamefont {Yoshida}}, \bibinfo {author}
  {\bibfnamefont {M.}~\bibnamefont {Kurooka}}, \ and\ \bibinfo {author}
  {\bibfnamefont {J.}~\bibnamefont {Mizuki}},\ }\bibfield  {title} {\enquote
  {\bibinfo {title} {{High-energy spin and charge excitations in electron-doped
  copper oxide superconductors}},}\ }\href
  {http://dx.doi.org/10.1038/ncomms4714} {\bibfield  {journal} {\bibinfo
  {journal} {Nat. Commun.}\ }\textbf {\bibinfo {volume} {5}},\ \bibinfo {pages}
  {3714} (\bibinfo {year} {2014})}\BibitemShut {NoStop}%
\bibitem [{\citenamefont {Lee}\ \emph {et~al.}(2014)\citenamefont {Lee},
  \citenamefont {Lee}, \citenamefont {Nowadnick}, \citenamefont {Gerber},
  \citenamefont {Tabi\'{s}}, \citenamefont {Huang}, \citenamefont {Strocov},
  \citenamefont {Motoyama}, \citenamefont {Yu}, \citenamefont {Moritz},
  \citenamefont {Huang}, \citenamefont {Wang}, \citenamefont {Huang},
  \citenamefont {Wu}, \citenamefont {Chen}, \citenamefont {Huang},
  \citenamefont {Greven}, \citenamefont {Schmitt}, \citenamefont {Shen},\ and\
  \citenamefont {Devereaux}}]{LeeNatPhys2014}%
  \BibitemOpen
  \bibfield  {author} {\bibinfo {author} {\bibfnamefont {W.~S.}\ \bibnamefont
  {Lee}}, \bibinfo {author} {\bibfnamefont {J.~J.}\ \bibnamefont {Lee}},
  \bibinfo {author} {\bibfnamefont {E.~A.}\ \bibnamefont {Nowadnick}}, \bibinfo
  {author} {\bibfnamefont {S.}~\bibnamefont {Gerber}}, \bibinfo {author}
  {\bibfnamefont {W.}~\bibnamefont {Tabi\'{s}}}, \bibinfo {author}
  {\bibfnamefont {S.~W.}\ \bibnamefont {Huang}}, \bibinfo {author}
  {\bibfnamefont {V.~N.}\ \bibnamefont {Strocov}}, \bibinfo {author}
  {\bibfnamefont {E.~M.}\ \bibnamefont {Motoyama}}, \bibinfo {author}
  {\bibfnamefont {G.}~\bibnamefont {Yu}}, \bibinfo {author} {\bibfnamefont
  {B.}~\bibnamefont {Moritz}}, \bibinfo {author} {\bibfnamefont {H.~Y.}\
  \bibnamefont {Huang}}, \bibinfo {author} {\bibfnamefont {R.~P.}\ \bibnamefont
  {Wang}}, \bibinfo {author} {\bibfnamefont {Y.~B.}\ \bibnamefont {Huang}},
  \bibinfo {author} {\bibfnamefont {W.~B.}\ \bibnamefont {Wu}}, \bibinfo
  {author} {\bibfnamefont {C.~T.}\ \bibnamefont {Chen}}, \bibinfo {author}
  {\bibfnamefont {D.~J.}\ \bibnamefont {Huang}}, \bibinfo {author}
  {\bibfnamefont {M.}~\bibnamefont {Greven}}, \bibinfo {author} {\bibfnamefont
  {T.}~\bibnamefont {Schmitt}}, \bibinfo {author} {\bibfnamefont {Z.~X.}\
  \bibnamefont {Shen}}, \ and\ \bibinfo {author} {\bibfnamefont {T.~P.}\
  \bibnamefont {Devereaux}},\ }\bibfield  {title} {\enquote {\bibinfo {title}
  {{Asymmetry of collective excitations in electron- and hole-doped cuprate
  superconductors}},}\ }\href {http://dx.doi.org/10.1038/nphys3117} {\bibfield
  {journal} {\bibinfo  {journal} {Nat. Phys.}\ }\textbf {\bibinfo {volume}
  {10}},\ \bibinfo {pages} {883} (\bibinfo {year} {2014})}\BibitemShut
  {NoStop}%
\bibitem [{\citenamefont {Guarise}\ \emph {et~al.}(2014)\citenamefont
  {Guarise}, \citenamefont {Piazza}, \citenamefont {Berger}, \citenamefont
  {Giannini}, \citenamefont {Schmitt}, \citenamefont {R{\o}nnow}, \citenamefont
  {Sawatzky}, \citenamefont {van~den Brink}, \citenamefont {Altenfeld},
  \citenamefont {Eremin},\ and\ \citenamefont {Grioni}}]{GuariseNatCommun2014}%
  \BibitemOpen
  \bibfield  {author} {\bibinfo {author} {\bibfnamefont {M.}~\bibnamefont
  {Guarise}}, \bibinfo {author} {\bibfnamefont {B.~Dalla}\ \bibnamefont
  {Piazza}}, \bibinfo {author} {\bibfnamefont {H.}~\bibnamefont {Berger}},
  \bibinfo {author} {\bibfnamefont {E.}~\bibnamefont {Giannini}}, \bibinfo
  {author} {\bibfnamefont {T.}~\bibnamefont {Schmitt}}, \bibinfo {author}
  {\bibfnamefont {H.~M.}\ \bibnamefont {R{\o}nnow}}, \bibinfo {author}
  {\bibfnamefont {G.~A.}\ \bibnamefont {Sawatzky}}, \bibinfo {author}
  {\bibfnamefont {J.}~\bibnamefont {van~den Brink}}, \bibinfo {author}
  {\bibfnamefont {D.}~\bibnamefont {Altenfeld}}, \bibinfo {author}
  {\bibfnamefont {I.}~\bibnamefont {Eremin}}, \ and\ \bibinfo {author}
  {\bibfnamefont {M.}~\bibnamefont {Grioni}},\ }\bibfield  {title} {\enquote
  {\bibinfo {title} {{Anisotropic softening of magnetic excitations along the
  nodal direction in superconducting cuprates}},}\ }\href
  {http://dx.doi.org/10.1038/ncomms6760} {\bibfield  {journal} {\bibinfo
  {journal} {Nat. Commun.}\ }\textbf {\bibinfo {volume} {5}},\ \bibinfo {pages}
  {5760} (\bibinfo {year} {2014})}\BibitemShut {NoStop}%
\bibitem [{\citenamefont {Wakimoto}\ \emph {et~al.}(2015)\citenamefont
  {Wakimoto}, \citenamefont {Ishii}, \citenamefont {Kimura}, \citenamefont
  {Fujita}, \citenamefont {Dellea}, \citenamefont {Kummer}, \citenamefont
  {Braicovich}, \citenamefont {Ghiringhelli}, \citenamefont {Debeer-Schmitt},\
  and\ \citenamefont {Granroth}}]{WakimotoPhysRevB2015}%
  \BibitemOpen
  \bibfield  {author} {\bibinfo {author} {\bibfnamefont {S.}~\bibnamefont
  {Wakimoto}}, \bibinfo {author} {\bibfnamefont {K.}~\bibnamefont {Ishii}},
  \bibinfo {author} {\bibfnamefont {H.}~\bibnamefont {Kimura}}, \bibinfo
  {author} {\bibfnamefont {M.}~\bibnamefont {Fujita}}, \bibinfo {author}
  {\bibfnamefont {G.}~\bibnamefont {Dellea}}, \bibinfo {author} {\bibfnamefont
  {K.}~\bibnamefont {Kummer}}, \bibinfo {author} {\bibfnamefont
  {L.}~\bibnamefont {Braicovich}}, \bibinfo {author} {\bibfnamefont
  {G.}~\bibnamefont {Ghiringhelli}}, \bibinfo {author} {\bibfnamefont {L.~M.}\
  \bibnamefont {Debeer-Schmitt}}, \ and\ \bibinfo {author} {\bibfnamefont
  {G.~E.}\ \bibnamefont {Granroth}},\ }\bibfield  {title} {\enquote {\bibinfo
  {title} {{High-energy magnetic excitations in overdoped
  $\mathrm{La_{2-\mathit{x}}Sr_\mathit{x}CuO_4}$ studied by neutron and
  resonant inelastic $x$-ray scattering}},}\ }\href
  {http://dx.doi.org/10.1103/physrevb.91.184513} {\bibfield  {journal}
  {\bibinfo  {journal} {Phys. Rev. B}\ }\textbf {\bibinfo {volume} {91}}
  (\bibinfo {year} {2015})}\BibitemShut {NoStop}%
\bibitem [{\citenamefont {Minola}\ \emph {et~al.}(2017)\citenamefont {Minola},
  \citenamefont {Lu}, \citenamefont {Peng}, \citenamefont {Dellea},
  \citenamefont {Gretarsson}, \citenamefont {Haverkort}, \citenamefont {Ding},
  \citenamefont {Sun}, \citenamefont {Zhou}, \citenamefont {Peets},
  \citenamefont {Chauviere}, \citenamefont {Dosanjh}, \citenamefont {Bonn},
  \citenamefont {Liang}, \citenamefont {Damascelli}, \citenamefont {Dantz},
  \citenamefont {Lu}, \citenamefont {Schmitt}, \citenamefont {Braicovich},
  \citenamefont {Ghiringhelli}, \citenamefont {Keimer},\ and\ \citenamefont
  {Le~Tacon}}]{MinolaPhysRevLett2017}%
  \BibitemOpen
  \bibfield  {author} {\bibinfo {author} {\bibfnamefont {M.}~\bibnamefont
  {Minola}}, \bibinfo {author} {\bibfnamefont {Y.}~\bibnamefont {Lu}}, \bibinfo
  {author} {\bibfnamefont {Y.~Y.}\ \bibnamefont {Peng}}, \bibinfo {author}
  {\bibfnamefont {G.}~\bibnamefont {Dellea}}, \bibinfo {author} {\bibfnamefont
  {H.}~\bibnamefont {Gretarsson}}, \bibinfo {author} {\bibfnamefont {M.~W.}\
  \bibnamefont {Haverkort}}, \bibinfo {author} {\bibfnamefont {Y.}~\bibnamefont
  {Ding}}, \bibinfo {author} {\bibfnamefont {X.}~\bibnamefont {Sun}}, \bibinfo
  {author} {\bibfnamefont {X.~J.}\ \bibnamefont {Zhou}}, \bibinfo {author}
  {\bibfnamefont {D.~C.}\ \bibnamefont {Peets}}, \bibinfo {author}
  {\bibfnamefont {L.}~\bibnamefont {Chauviere}}, \bibinfo {author}
  {\bibfnamefont {P.}~\bibnamefont {Dosanjh}}, \bibinfo {author} {\bibfnamefont
  {D.~A.}\ \bibnamefont {Bonn}}, \bibinfo {author} {\bibfnamefont
  {R.}~\bibnamefont {Liang}}, \bibinfo {author} {\bibfnamefont
  {A.}~\bibnamefont {Damascelli}}, \bibinfo {author} {\bibfnamefont
  {M.}~\bibnamefont {Dantz}}, \bibinfo {author} {\bibfnamefont
  {X.}~\bibnamefont {Lu}}, \bibinfo {author} {\bibfnamefont {T.}~\bibnamefont
  {Schmitt}}, \bibinfo {author} {\bibfnamefont {L.}~\bibnamefont {Braicovich}},
  \bibinfo {author} {\bibfnamefont {G.}~\bibnamefont {Ghiringhelli}}, \bibinfo
  {author} {\bibfnamefont {B.}~\bibnamefont {Keimer}}, \ and\ \bibinfo {author}
  {\bibfnamefont {M.}~\bibnamefont {Le~Tacon}},\ }\bibfield  {title} {\enquote
  {\bibinfo {title} {{Crossover from Collective to Incoherent Spin Excitations
  in Superconducting Cuprates Probed by Detuned Resonant Inelastic $X$-Ray
  Scattering}},}\ }\href {http://dx.doi.org/10.1103/physrevlett.119.097001}
  {\bibfield  {journal} {\bibinfo  {journal} {Phys. Rev. Lett.}\ }\textbf
  {\bibinfo {volume} {119}},\ \bibinfo {pages} {245133} (\bibinfo {year}
  {2017})}\BibitemShut {NoStop}%
\bibitem [{\citenamefont {Ivashko}\ \emph {et~al.}(2017)\citenamefont
  {Ivashko}, \citenamefont {Shaik}, \citenamefont {Lu}, \citenamefont
  {Fatuzzo}, \citenamefont {Dantz}, \citenamefont {Freeman}, \citenamefont
  {McNally}, \citenamefont {Destraz}, \citenamefont {Christensen},
  \citenamefont {Kurosawa}, \citenamefont {Momono}, \citenamefont {Oda},
  \citenamefont {Matt}, \citenamefont {Monney}, \citenamefont {R\o{}nnow},
  \citenamefont {Schmitt},\ and\ \citenamefont {Chang}}]{IvashkoPhysRevB2017}%
  \BibitemOpen
  \bibfield  {author} {\bibinfo {author} {\bibfnamefont {O.}~\bibnamefont
  {Ivashko}}, \bibinfo {author} {\bibfnamefont {N.~E.}\ \bibnamefont {Shaik}},
  \bibinfo {author} {\bibfnamefont {X.}~\bibnamefont {Lu}}, \bibinfo {author}
  {\bibfnamefont {C.~G.}\ \bibnamefont {Fatuzzo}}, \bibinfo {author}
  {\bibfnamefont {M.}~\bibnamefont {Dantz}}, \bibinfo {author} {\bibfnamefont
  {P.~G.}\ \bibnamefont {Freeman}}, \bibinfo {author} {\bibfnamefont {D.~E.}\
  \bibnamefont {McNally}}, \bibinfo {author} {\bibfnamefont {D.}~\bibnamefont
  {Destraz}}, \bibinfo {author} {\bibfnamefont {N.~B.}\ \bibnamefont
  {Christensen}}, \bibinfo {author} {\bibfnamefont {T.}~\bibnamefont
  {Kurosawa}}, \bibinfo {author} {\bibfnamefont {N.}~\bibnamefont {Momono}},
  \bibinfo {author} {\bibfnamefont {M.}~\bibnamefont {Oda}}, \bibinfo {author}
  {\bibfnamefont {C.~E.}\ \bibnamefont {Matt}}, \bibinfo {author}
  {\bibfnamefont {C.}~\bibnamefont {Monney}}, \bibinfo {author} {\bibfnamefont
  {H.~M.}\ \bibnamefont {R\o{}nnow}}, \bibinfo {author} {\bibfnamefont
  {T.}~\bibnamefont {Schmitt}}, \ and\ \bibinfo {author} {\bibfnamefont
  {J.}~\bibnamefont {Chang}},\ }\bibfield  {title} {\enquote {\bibinfo {title}
  {{Damped spin excitations in a doped cuprate superconductor with orbital
  hybridization}},}\ }\href {http://dx.doi.org/10.1103/physrevb.95.214508}
  {\bibfield  {journal} {\bibinfo  {journal} {Phys. Rev. B}\ }\textbf {\bibinfo
  {volume} {95}} (\bibinfo {year} {2017})}\BibitemShut {NoStop}%
\bibitem [{\citenamefont {Meyers}\ \emph {et~al.}(2017)\citenamefont {Meyers},
  \citenamefont {Miao}, \citenamefont {Walters}, \citenamefont {Bisogni},
  \citenamefont {Springell}, \citenamefont {d'Astuto}, \citenamefont {Dantz},
  \citenamefont {Pelliciari}, \citenamefont {Huang}, \citenamefont {Okamoto},
  \citenamefont {Huang}, \citenamefont {Hill}, \citenamefont {He},
  \citenamefont {Bo\ifmmode \check{z}\else \v{z}\fi{}ovi\ifmmode~\acute{c}\else
  \'{c}\fi{}}, \citenamefont {Schmitt},\ and\ \citenamefont
  {Dean}}]{MeyersPhysRevB2017}%
  \BibitemOpen
  \bibfield  {author} {\bibinfo {author} {\bibfnamefont {D.}~\bibnamefont
  {Meyers}}, \bibinfo {author} {\bibfnamefont {H.}~\bibnamefont {Miao}},
  \bibinfo {author} {\bibfnamefont {A.~C.}\ \bibnamefont {Walters}}, \bibinfo
  {author} {\bibfnamefont {V.}~\bibnamefont {Bisogni}}, \bibinfo {author}
  {\bibfnamefont {R.~S.}\ \bibnamefont {Springell}}, \bibinfo {author}
  {\bibfnamefont {M.}~\bibnamefont {d'Astuto}}, \bibinfo {author}
  {\bibfnamefont {M.}~\bibnamefont {Dantz}}, \bibinfo {author} {\bibfnamefont
  {J.}~\bibnamefont {Pelliciari}}, \bibinfo {author} {\bibfnamefont {H.~Y.}\
  \bibnamefont {Huang}}, \bibinfo {author} {\bibfnamefont {J.}~\bibnamefont
  {Okamoto}}, \bibinfo {author} {\bibfnamefont {D.~J.}\ \bibnamefont {Huang}},
  \bibinfo {author} {\bibfnamefont {J.~P.}\ \bibnamefont {Hill}}, \bibinfo
  {author} {\bibfnamefont {X.}~\bibnamefont {He}}, \bibinfo {author}
  {\bibfnamefont {I.}~\bibnamefont {Bo\ifmmode \check{z}\else
  \v{z}\fi{}ovi\ifmmode~\acute{c}\else \'{c}\fi{}}}, \bibinfo {author}
  {\bibfnamefont {T.}~\bibnamefont {Schmitt}}, \ and\ \bibinfo {author}
  {\bibfnamefont {M.~P.~M.}\ \bibnamefont {Dean}},\ }\bibfield  {title}
  {\enquote {\bibinfo {title} {{Doping dependence of the magnetic excitations
  in $\mathrm{La_{2-\mathit{x}}Sr_\mathit{x}CuO_4}$}},}\ }\href
  {http://dx.doi.org/10.1103/PhysRevB.95.075139} {\bibfield  {journal}
  {\bibinfo  {journal} {Phys. Rev. B}\ }\textbf {\bibinfo {volume} {95}},\
  \bibinfo {pages} {075139} (\bibinfo {year} {2017})}\BibitemShut {NoStop}%
\bibitem [{\citenamefont {L.~Chaix}\ and\ \citenamefont
  {Lee}(2018)}]{ChaixPhysRevB2018}%
  \BibitemOpen
  \bibfield  {author} {\bibinfo {author} {\bibfnamefont {S.~Gerber X. Lu C. Jia
  Y. Huang D. E. McNally Y. Wang F. H. Vernay A. Keren M. Shi B. Moritz Z.-X.
  Shen T. Schmitt T. P.~Devereaux}\ \bibnamefont {L.~Chaix}, \bibfnamefont
  {E.~W.~Huang}}\ and\ \bibinfo {author} {\bibfnamefont {W.-S.}\ \bibnamefont
  {Lee}},\ }\bibfield  {title} {\enquote {\bibinfo {title} {{Resonant inelastic
  $x$-ray scattering studies of magnons band bimagnons in the lightly doped
  cuprate $\mathrm{La_{2-\mathit{x}}Sr_\mathit{x}CuO_4}$}},}\ }\href
  {http://dx.doi.org/10.1103/physrevb.97.155144} {\bibfield  {journal}
  {\bibinfo  {journal} {Phys. Rev. B}\ }\textbf {\bibinfo {volume} {97}},\
  \bibinfo {pages} {155144} (\bibinfo {year} {2018})}\BibitemShut {NoStop}%
\bibitem [{\citenamefont {Robarts}\ \emph {et~al.}(2019)\citenamefont
  {Robarts}, \citenamefont {Barth\'elemy}, \citenamefont {Kummer},
  \citenamefont {Garc\'{\i}a-Fern\'andez}, \citenamefont {Li}, \citenamefont
  {Nag}, \citenamefont {Walters}, \citenamefont {Zhou},\ and\ \citenamefont
  {Hayden}}]{Robarts_arXiV_2019}%
  \BibitemOpen
  \bibfield  {author} {\bibinfo {author} {\bibfnamefont {H.~C.}\ \bibnamefont
  {Robarts}}, \bibinfo {author} {\bibfnamefont {M.}~\bibnamefont
  {Barth\'elemy}}, \bibinfo {author} {\bibfnamefont {K.}~\bibnamefont
  {Kummer}}, \bibinfo {author} {\bibfnamefont {M.}~\bibnamefont
  {Garc\'{\i}a-Fern\'andez}}, \bibinfo {author} {\bibfnamefont
  {J.}~\bibnamefont {Li}}, \bibinfo {author} {\bibfnamefont {A.}~\bibnamefont
  {Nag}}, \bibinfo {author} {\bibfnamefont {A.~C.}\ \bibnamefont {Walters}},
  \bibinfo {author} {\bibfnamefont {K.~J.}\ \bibnamefont {Zhou}}, \ and\
  \bibinfo {author} {\bibfnamefont {S.~M.}\ \bibnamefont {Hayden}},\ }\bibfield
   {title} {\enquote {\bibinfo {title} {{Anisotropic damping and wave vector
  dependent susceptibility of the spin fluctuations in
  ${\mathrm{La}}_{2\ensuremath{-}x}{\mathrm{Sr}}_{x}{\mathrm{CuO}}_{4}$ studied
  by resonant inelastic x-ray scattering}},}\ }\href {\doibase
  10.1103/PhysRevB.100.214510} {\bibfield  {journal} {\bibinfo  {journal}
  {Phys. Rev. B}\ }\textbf {\bibinfo {volume} {100}},\ \bibinfo {pages}
  {214510} (\bibinfo {year} {2019})}\BibitemShut {NoStop}%
\bibitem [{\citenamefont {Zhou}\ \emph {et~al.}(2013)\citenamefont {Zhou},
  \citenamefont {Huang}, \citenamefont {Monney}, \citenamefont {Dai},
  \citenamefont {Strocov}, \citenamefont {Wang}, \citenamefont {Chen},
  \citenamefont {Zhang}, \citenamefont {Dai}, \citenamefont {Patthey},
  \citenamefont {van~den Brink}, \citenamefont {Ding},\ and\ \citenamefont
  {Schmitt}}]{ZhouNatCommun2013}%
  \BibitemOpen
  \bibfield  {author} {\bibinfo {author} {\bibfnamefont {K.-J.}\ \bibnamefont
  {Zhou}}, \bibinfo {author} {\bibfnamefont {Y.-B.}\ \bibnamefont {Huang}},
  \bibinfo {author} {\bibfnamefont {C.}~\bibnamefont {Monney}}, \bibinfo
  {author} {\bibfnamefont {X.}~\bibnamefont {Dai}}, \bibinfo {author}
  {\bibfnamefont {V.~N.}\ \bibnamefont {Strocov}}, \bibinfo {author}
  {\bibfnamefont {N.-L.}\ \bibnamefont {Wang}}, \bibinfo {author}
  {\bibfnamefont {Z.-G.}\ \bibnamefont {Chen}}, \bibinfo {author}
  {\bibfnamefont {C.}~\bibnamefont {Zhang}}, \bibinfo {author} {\bibfnamefont
  {P.}~\bibnamefont {Dai}}, \bibinfo {author} {\bibfnamefont {L.}~\bibnamefont
  {Patthey}}, \bibinfo {author} {\bibfnamefont {J.}~\bibnamefont {van~den
  Brink}}, \bibinfo {author} {\bibfnamefont {H.}~\bibnamefont {Ding}}, \ and\
  \bibinfo {author} {\bibfnamefont {T.}~\bibnamefont {Schmitt}},\ }\bibfield
  {title} {\enquote {\bibinfo {title} {{Persistent high-energy spin excitations
  in iron-pnictide superconductors}},}\ }\href
  {http://dx.doi.org/10.1038/ncomms2428} {\bibfield  {journal} {\bibinfo
  {journal} {Nat. Commun.}\ }\textbf {\bibinfo {volume} {4}},\ \bibinfo {pages}
  {1470} (\bibinfo {year} {2013})}\BibitemShut {NoStop}%
\bibitem [{\citenamefont {Gretarsson}\ \emph {et~al.}(2016)\citenamefont
  {Gretarsson}, \citenamefont {Sung}, \citenamefont {Porras}, \citenamefont
  {Bertinshaw}, \citenamefont {Dietl}, \citenamefont {Bruin}, \citenamefont
  {Bangura}, \citenamefont {Kim}, \citenamefont {Dinnebier}, \citenamefont
  {Kim}, \citenamefont {Al-Zein}, \citenamefont {Moretti~Sala}, \citenamefont
  {Krisch}, \citenamefont {Le~Tacon}, \citenamefont {Keimer},\ and\
  \citenamefont {Kim}}]{GretarssonPhysRevLett2016}%
  \BibitemOpen
  \bibfield  {author} {\bibinfo {author} {\bibfnamefont {H.}~\bibnamefont
  {Gretarsson}}, \bibinfo {author} {\bibfnamefont {N.~H.}\ \bibnamefont
  {Sung}}, \bibinfo {author} {\bibfnamefont {J.}~\bibnamefont {Porras}},
  \bibinfo {author} {\bibfnamefont {J.}~\bibnamefont {Bertinshaw}}, \bibinfo
  {author} {\bibfnamefont {C.}~\bibnamefont {Dietl}}, \bibinfo {author}
  {\bibfnamefont {Jan A.~N.}\ \bibnamefont {Bruin}}, \bibinfo {author}
  {\bibfnamefont {A.~F.}\ \bibnamefont {Bangura}}, \bibinfo {author}
  {\bibfnamefont {Y.~K.}\ \bibnamefont {Kim}}, \bibinfo {author} {\bibfnamefont
  {R.}~\bibnamefont {Dinnebier}}, \bibinfo {author} {\bibfnamefont {Jungho}\
  \bibnamefont {Kim}}, \bibinfo {author} {\bibfnamefont {A.}~\bibnamefont
  {Al-Zein}}, \bibinfo {author} {\bibfnamefont {M.}~\bibnamefont
  {Moretti~Sala}}, \bibinfo {author} {\bibfnamefont {M.}~\bibnamefont
  {Krisch}}, \bibinfo {author} {\bibfnamefont {M.}~\bibnamefont {Le~Tacon}},
  \bibinfo {author} {\bibfnamefont {B.}~\bibnamefont {Keimer}}, \ and\ \bibinfo
  {author} {\bibfnamefont {B.~J.}\ \bibnamefont {Kim}},\ }\bibfield  {title}
  {\enquote {\bibinfo {title} {{Persistent Paramagnons Deep in the Metallic
  Phase of $\mathrm{Sr_{2-\mathit{x}}La_\mathit{x}IrO_4}$}},}\ }\href
  {http://dx.doi.org/10.1103/PhysRevLett.117.107001} {\bibfield  {journal}
  {\bibinfo  {journal} {Phys. Rev. Lett.}\ }\textbf {\bibinfo {volume} {117}},\
  \bibinfo {pages} {107001} (\bibinfo {year} {2016})}\BibitemShut {NoStop}%
\bibitem [{\citenamefont {Fumagalli}\ \emph {et~al.}(2019)\citenamefont
  {Fumagalli}, \citenamefont {Braicovich}, \citenamefont {Minola},
  \citenamefont {Peng}, \citenamefont {Kummer}, \citenamefont {Betto},
  \citenamefont {Rossi}, \citenamefont {Lefran\ifmmode~\mbox{\c{c}}\else
  \c{c}\fi{}ois}, \citenamefont {Morawe}, \citenamefont {Salluzzo},
  \citenamefont {Suzuki}, \citenamefont {Yakhou}, \citenamefont {Le~Tacon},
  \citenamefont {Keimer}, \citenamefont {Brookes}, \citenamefont {Sala},\ and\
  \citenamefont {Ghiringhelli}}]{FumagalliPhysRevB2019}%
  \BibitemOpen
  \bibfield  {author} {\bibinfo {author} {\bibfnamefont {R.}~\bibnamefont
  {Fumagalli}}, \bibinfo {author} {\bibfnamefont {L.}~\bibnamefont
  {Braicovich}}, \bibinfo {author} {\bibfnamefont {M.}~\bibnamefont {Minola}},
  \bibinfo {author} {\bibfnamefont {Y.~Y.}\ \bibnamefont {Peng}}, \bibinfo
  {author} {\bibfnamefont {K.}~\bibnamefont {Kummer}}, \bibinfo {author}
  {\bibfnamefont {D.}~\bibnamefont {Betto}}, \bibinfo {author} {\bibfnamefont
  {M.}~\bibnamefont {Rossi}}, \bibinfo {author} {\bibfnamefont
  {E.}~\bibnamefont {Lefran\ifmmode~\mbox{\c{c}}\else \c{c}\fi{}ois}}, \bibinfo
  {author} {\bibfnamefont {C.}~\bibnamefont {Morawe}}, \bibinfo {author}
  {\bibfnamefont {M.}~\bibnamefont {Salluzzo}}, \bibinfo {author}
  {\bibfnamefont {H.}~\bibnamefont {Suzuki}}, \bibinfo {author} {\bibfnamefont
  {F.}~\bibnamefont {Yakhou}}, \bibinfo {author} {\bibfnamefont
  {M.}~\bibnamefont {Le~Tacon}}, \bibinfo {author} {\bibfnamefont
  {B.}~\bibnamefont {Keimer}}, \bibinfo {author} {\bibfnamefont {N.~B.}\
  \bibnamefont {Brookes}}, \bibinfo {author} {\bibfnamefont {M.~Moretti}\
  \bibnamefont {Sala}}, \ and\ \bibinfo {author} {\bibfnamefont
  {G.}~\bibnamefont {Ghiringhelli}},\ }\bibfield  {title} {\enquote {\bibinfo
  {title} {{Polarization-resolved Cu $L_3$-edge resonant inelastic $x$-ray
  scattering of orbital and spin excitations in
  $\mathrm{NdBa_2Cu_3O_{7-\delta}}$}},}\ }\href {\doibase
  10.1103/physrevb.99.134517} {\bibfield  {journal} {\bibinfo  {journal} {Phys.
  Rev. B}\ }\textbf {\bibinfo {volume} {99}},\ \bibinfo {pages} {134517}
  (\bibinfo {year} {2019})}\BibitemShut {NoStop}%
\bibitem [{\citenamefont {Le~Tacon}\ \emph {et~al.}(2011)\citenamefont
  {Le~Tacon}, \citenamefont {Ghiringhelli}, \citenamefont {Chaloupka},
  \citenamefont {Sala}, \citenamefont {Hinkov}, \citenamefont {Haverkort},
  \citenamefont {Minola}, \citenamefont {Bakr}, \citenamefont {Zhou},
  \citenamefont {Blanco-Canosa}, \citenamefont {Monney}, \citenamefont {Song},
  \citenamefont {Sun}, \citenamefont {Lin}, \citenamefont {De~Luca},
  \citenamefont {Salluzzo}, \citenamefont {Khaliullin}, \citenamefont
  {Schmitt}, \citenamefont {Braicovich},\ and\ \citenamefont
  {Keimer}}]{LeTaconNatPhys2011}%
  \BibitemOpen
  \bibfield  {author} {\bibinfo {author} {\bibfnamefont {M.}~\bibnamefont
  {Le~Tacon}}, \bibinfo {author} {\bibfnamefont {G.}~\bibnamefont
  {Ghiringhelli}}, \bibinfo {author} {\bibfnamefont {J.}~\bibnamefont
  {Chaloupka}}, \bibinfo {author} {\bibfnamefont {M.~Moretti}\ \bibnamefont
  {Sala}}, \bibinfo {author} {\bibfnamefont {V.}~\bibnamefont {Hinkov}},
  \bibinfo {author} {\bibfnamefont {M.~W.}\ \bibnamefont {Haverkort}}, \bibinfo
  {author} {\bibfnamefont {M.}~\bibnamefont {Minola}}, \bibinfo {author}
  {\bibfnamefont {M.}~\bibnamefont {Bakr}}, \bibinfo {author} {\bibfnamefont
  {K.~J.}\ \bibnamefont {Zhou}}, \bibinfo {author} {\bibfnamefont
  {S.}~\bibnamefont {Blanco-Canosa}}, \bibinfo {author} {\bibfnamefont
  {C.}~\bibnamefont {Monney}}, \bibinfo {author} {\bibfnamefont {Y.~T.}\
  \bibnamefont {Song}}, \bibinfo {author} {\bibfnamefont {G.~L.}\ \bibnamefont
  {Sun}}, \bibinfo {author} {\bibfnamefont {C.~T.}\ \bibnamefont {Lin}},
  \bibinfo {author} {\bibfnamefont {G.~M.}\ \bibnamefont {De~Luca}}, \bibinfo
  {author} {\bibfnamefont {M.}~\bibnamefont {Salluzzo}}, \bibinfo {author}
  {\bibfnamefont {G.}~\bibnamefont {Khaliullin}}, \bibinfo {author}
  {\bibfnamefont {T.}~\bibnamefont {Schmitt}}, \bibinfo {author} {\bibfnamefont
  {L.}~\bibnamefont {Braicovich}}, \ and\ \bibinfo {author} {\bibfnamefont
  {B.}~\bibnamefont {Keimer}},\ }\bibfield  {title} {\enquote {\bibinfo {title}
  {{Intense paramagnon excitations in a large family of high-temperature
  superconductors}},}\ }\href {http://dx.doi.org/10.1038/nphys2041} {\bibfield
  {journal} {\bibinfo  {journal} {Nat. Phys.}\ }\textbf {\bibinfo {volume}
  {7}},\ \bibinfo {pages} {725} (\bibinfo {year} {2011})}\BibitemShut {NoStop}%
\bibitem [{\citenamefont {Jia}\ \emph {et~al.}(2014)\citenamefont {Jia},
  \citenamefont {Nowadnick}, \citenamefont {Wohlfeld}, \citenamefont {Kung},
  \citenamefont {Chen}, \citenamefont {Johnston}, \citenamefont {Tohyama},
  \citenamefont {Moritz},\ and\ \citenamefont {Devereaux}}]{JiaNatCommun2014}%
  \BibitemOpen
  \bibfield  {author} {\bibinfo {author} {\bibfnamefont {C.~J.}\ \bibnamefont
  {Jia}}, \bibinfo {author} {\bibfnamefont {E.~A.}\ \bibnamefont {Nowadnick}},
  \bibinfo {author} {\bibfnamefont {K.}~\bibnamefont {Wohlfeld}}, \bibinfo
  {author} {\bibfnamefont {Y.~F.}\ \bibnamefont {Kung}}, \bibinfo {author}
  {\bibfnamefont {C.-C.}\ \bibnamefont {Chen}}, \bibinfo {author}
  {\bibfnamefont {S.}~\bibnamefont {Johnston}}, \bibinfo {author}
  {\bibfnamefont {T.}~\bibnamefont {Tohyama}}, \bibinfo {author} {\bibfnamefont
  {B.}~\bibnamefont {Moritz}}, \ and\ \bibinfo {author} {\bibfnamefont {T.~P.}\
  \bibnamefont {Devereaux}},\ }\bibfield  {title} {\enquote {\bibinfo {title}
  {{Persistent spin excitations in doped antiferromagnets revealed by resonant
  inelastic light scattering}},}\ }\href {http://dx.doi.org/10.1038/ncomms4314}
  {\bibfield  {journal} {\bibinfo  {journal} {Nat. Commun.}\ }\textbf {\bibinfo
  {volume} {5}},\ \bibinfo {pages} {3314} (\bibinfo {year} {2014})}\BibitemShut
  {NoStop}%
\bibitem [{\citenamefont {Peng}\ \emph {et~al.}(2018)\citenamefont {Peng},
  \citenamefont {Huang}, \citenamefont {Fumagalli}, \citenamefont {Minola},
  \citenamefont {Wang}, \citenamefont {Sun}, \citenamefont {Ding},
  \citenamefont {Kummer}, \citenamefont {Zhou}, \citenamefont {Brookes},
  \citenamefont {Moritz}, \citenamefont {Braicovich}, \citenamefont
  {Devereaux},\ and\ \citenamefont {Ghiringhelli}}]{PengPhysRevB2018}%
  \BibitemOpen
  \bibfield  {author} {\bibinfo {author} {\bibfnamefont {Y.~Y.}\ \bibnamefont
  {Peng}}, \bibinfo {author} {\bibfnamefont {E.~W.}\ \bibnamefont {Huang}},
  \bibinfo {author} {\bibfnamefont {R.}~\bibnamefont {Fumagalli}}, \bibinfo
  {author} {\bibfnamefont {M.}~\bibnamefont {Minola}}, \bibinfo {author}
  {\bibfnamefont {Y.}~\bibnamefont {Wang}}, \bibinfo {author} {\bibfnamefont
  {X.}~\bibnamefont {Sun}}, \bibinfo {author} {\bibfnamefont {Y.}~\bibnamefont
  {Ding}}, \bibinfo {author} {\bibfnamefont {K.}~\bibnamefont {Kummer}},
  \bibinfo {author} {\bibfnamefont {X.~J.}\ \bibnamefont {Zhou}}, \bibinfo
  {author} {\bibfnamefont {N.~B.}\ \bibnamefont {Brookes}}, \bibinfo {author}
  {\bibfnamefont {B.}~\bibnamefont {Moritz}}, \bibinfo {author} {\bibfnamefont
  {L.}~\bibnamefont {Braicovich}}, \bibinfo {author} {\bibfnamefont {T.~P.}\
  \bibnamefont {Devereaux}}, \ and\ \bibinfo {author} {\bibfnamefont
  {G.}~\bibnamefont {Ghiringhelli}},\ }\bibfield  {title} {\enquote {\bibinfo
  {title} {{Dispersion, damping, and intensity of spin excitations in the
  monolayer $\mathrm{(Bi,Pb)_2(Sr,La)_2CuO_{6+\delta}}$ cuprate superconductor
  family}},}\ }\href {http://dx.doi.org/10.1103/physrevb.98.144507} {\bibfield
  {journal} {\bibinfo  {journal} {Phys. Rev. B}\ }\textbf {\bibinfo {volume}
  {98}} (\bibinfo {year} {2018})}\BibitemShut {NoStop}%
\bibitem [{\citenamefont {Ishii}\ \emph {et~al.}(2017)\citenamefont {Ishii},
  \citenamefont {Tohyama}, \citenamefont {Asano}, \citenamefont {Sato},
  \citenamefont {Fujita}, \citenamefont {Wakimoto}, \citenamefont {Tustsui},
  \citenamefont {Sota}, \citenamefont {Miyawaki}, \citenamefont {Niwa},
  \citenamefont {Harada}, \citenamefont {Pelliciari}, \citenamefont {Huang},
  \citenamefont {Schmitt}, \citenamefont {Yamamoto},\ and\ \citenamefont
  {Mizuki}}]{IshiiPhysRevB2017}%
  \BibitemOpen
  \bibfield  {author} {\bibinfo {author} {\bibfnamefont {K.}~\bibnamefont
  {Ishii}}, \bibinfo {author} {\bibfnamefont {T.}~\bibnamefont {Tohyama}},
  \bibinfo {author} {\bibfnamefont {S.}~\bibnamefont {Asano}}, \bibinfo
  {author} {\bibfnamefont {K.}~\bibnamefont {Sato}}, \bibinfo {author}
  {\bibfnamefont {M.}~\bibnamefont {Fujita}}, \bibinfo {author} {\bibfnamefont
  {S.}~\bibnamefont {Wakimoto}}, \bibinfo {author} {\bibfnamefont
  {K.}~\bibnamefont {Tustsui}}, \bibinfo {author} {\bibfnamefont
  {S.}~\bibnamefont {Sota}}, \bibinfo {author} {\bibfnamefont {J.}~\bibnamefont
  {Miyawaki}}, \bibinfo {author} {\bibfnamefont {H.}~\bibnamefont {Niwa}},
  \bibinfo {author} {\bibfnamefont {Y.}~\bibnamefont {Harada}}, \bibinfo
  {author} {\bibfnamefont {J.}~\bibnamefont {Pelliciari}}, \bibinfo {author}
  {\bibfnamefont {Y.}~\bibnamefont {Huang}}, \bibinfo {author} {\bibfnamefont
  {T.}~\bibnamefont {Schmitt}}, \bibinfo {author} {\bibfnamefont
  {Y.}~\bibnamefont {Yamamoto}}, \ and\ \bibinfo {author} {\bibfnamefont
  {J.}~\bibnamefont {Mizuki}},\ }\bibfield  {title} {\enquote {\bibinfo {title}
  {{Observation of momentum-dependent charge excitations in hole-doped cuprates
  using resonant inelastic $x$-ray scattering at the oxygen $K$ edge}},}\
  }\href {http://dx.doi.org/10.1103/physrevb.96.115148} {\bibfield  {journal}
  {\bibinfo  {journal} {Phys. Rev. B}\ }\textbf {\bibinfo {volume} {96}},\
  \bibinfo {pages} {115148} (\bibinfo {year} {2017})}\BibitemShut {NoStop}%
\bibitem [{\citenamefont {Hepting}\ \emph {et~al.}(2018)\citenamefont
  {Hepting}, \citenamefont {Chaix}, \citenamefont {Huang}, \citenamefont
  {Fumagalli}, \citenamefont {Peng}, \citenamefont {Moritz}, \citenamefont
  {Kummer}, \citenamefont {Brookes}, \citenamefont {Lee}, \citenamefont
  {Hashimoto}, \citenamefont {Sarkar}, \citenamefont {He}, \citenamefont
  {Rotundu}, \citenamefont {Lee}, \citenamefont {Greene}, \citenamefont
  {Braicovich}, \citenamefont {Ghiringhelli}, \citenamefont {Shen},
  \citenamefont {Devereaux},\ and\ \citenamefont {Lee}}]{HeptingNature2018}%
  \BibitemOpen
  \bibfield  {author} {\bibinfo {author} {\bibfnamefont {M.}~\bibnamefont
  {Hepting}}, \bibinfo {author} {\bibfnamefont {L.}~\bibnamefont {Chaix}},
  \bibinfo {author} {\bibfnamefont {E.~W.}\ \bibnamefont {Huang}}, \bibinfo
  {author} {\bibfnamefont {R.}~\bibnamefont {Fumagalli}}, \bibinfo {author}
  {\bibfnamefont {Y.~Y.}\ \bibnamefont {Peng}}, \bibinfo {author}
  {\bibfnamefont {B.}~\bibnamefont {Moritz}}, \bibinfo {author} {\bibfnamefont
  {K.}~\bibnamefont {Kummer}}, \bibinfo {author} {\bibfnamefont {N.~B.}\
  \bibnamefont {Brookes}}, \bibinfo {author} {\bibfnamefont {W.~C.}\
  \bibnamefont {Lee}}, \bibinfo {author} {\bibfnamefont {M.}~\bibnamefont
  {Hashimoto}}, \bibinfo {author} {\bibfnamefont {T.}~\bibnamefont {Sarkar}},
  \bibinfo {author} {\bibfnamefont {J.-F.}\ \bibnamefont {He}}, \bibinfo
  {author} {\bibfnamefont {C.~R.}\ \bibnamefont {Rotundu}}, \bibinfo {author}
  {\bibfnamefont {Y.~S.}\ \bibnamefont {Lee}}, \bibinfo {author} {\bibfnamefont
  {R.~L.}\ \bibnamefont {Greene}}, \bibinfo {author} {\bibfnamefont
  {L.}~\bibnamefont {Braicovich}}, \bibinfo {author} {\bibfnamefont
  {G.}~\bibnamefont {Ghiringhelli}}, \bibinfo {author} {\bibfnamefont {Z.~X.}\
  \bibnamefont {Shen}}, \bibinfo {author} {\bibfnamefont {T.~P.}\ \bibnamefont
  {Devereaux}}, \ and\ \bibinfo {author} {\bibfnamefont {W.~S.}\ \bibnamefont
  {Lee}},\ }\bibfield  {title} {\enquote {\bibinfo {title} {{Three-dimensional
  collective charge excitations in electron-doped copper oxide
  superconductors}},}\ }\href {http://dx.doi.org/10.1038/s41586-018-0648-3}
  {\bibfield  {journal} {\bibinfo  {journal} {Nature}\ }\textbf {\bibinfo
  {volume} {563}},\ \bibinfo {pages} {374} (\bibinfo {year}
  {2018})}\BibitemShut {NoStop}%
\bibitem [{\citenamefont {Ishii}\ \emph {et~al.}(2019)\citenamefont {Ishii},
  \citenamefont {Kurooka}, \citenamefont {Shimizu}, \citenamefont {Fujita},
  \citenamefont {Yamada},\ and\ \citenamefont
  {Mizuki}}]{IshiiJPhysSocJapan2019}%
  \BibitemOpen
  \bibfield  {author} {\bibinfo {author} {\bibfnamefont {K.}~\bibnamefont
  {Ishii}}, \bibinfo {author} {\bibfnamefont {M.}~\bibnamefont {Kurooka}},
  \bibinfo {author} {\bibfnamefont {Y.}~\bibnamefont {Shimizu}}, \bibinfo
  {author} {\bibfnamefont {M.}~\bibnamefont {Fujita}}, \bibinfo {author}
  {\bibfnamefont {K.}~\bibnamefont {Yamada}}, \ and\ \bibinfo {author}
  {\bibfnamefont {J.}~\bibnamefont {Mizuki}},\ }\bibfield  {title} {\enquote
  {\bibinfo {title} {{Charge Excitations in
  $\mathrm{Nd_{2-\mathit{x}}Ce_\mathit{x}CuO_4}$ Observed with Resonant
  Inelastic $X$-ray Scattering: Comparison of Cu $K$-edge with Cu
  $L_3$-edge}},}\ }\href {\doibase 10.7566/JPSJ.88.075001} {\bibfield
  {journal} {\bibinfo  {journal} {J. Phys. Soc. Japan}\ }\textbf {\bibinfo
  {volume} {88}},\ \bibinfo {pages} {075001} (\bibinfo {year}
  {2019})}\BibitemShut {NoStop}%
\bibitem [{\citenamefont {Lin}\ \emph {et~al.}(2020)\citenamefont {Lin},
  \citenamefont {Yuan}, \citenamefont {Jin}, \citenamefont {Yin}, \citenamefont
  {Li}, \citenamefont {Zhou}, \citenamefont {Lu}, \citenamefont {Dantz},
  \citenamefont {Schmitt}, \citenamefont {Ding}, \citenamefont {Guo},
  \citenamefont {P.~M.~Dean},\ and\ \citenamefont
  {Liu}}]{LinNPJQuantMater2020}%
  \BibitemOpen
  \bibfield  {author} {\bibinfo {author} {\bibfnamefont {J.}~\bibnamefont
  {Lin}}, \bibinfo {author} {\bibfnamefont {J.}~\bibnamefont {Yuan}}, \bibinfo
  {author} {\bibfnamefont {K.}~\bibnamefont {Jin}}, \bibinfo {author}
  {\bibfnamefont {Z.}~\bibnamefont {Yin}}, \bibinfo {author} {\bibfnamefont
  {Gang}\ \bibnamefont {Li}}, \bibinfo {author} {\bibfnamefont {K.-J.}\
  \bibnamefont {Zhou}}, \bibinfo {author} {\bibfnamefont {X.}~\bibnamefont
  {Lu}}, \bibinfo {author} {\bibfnamefont {M.}~\bibnamefont {Dantz}}, \bibinfo
  {author} {\bibfnamefont {T.}~\bibnamefont {Schmitt}}, \bibinfo {author}
  {\bibfnamefont {H.}~\bibnamefont {Ding}}, \bibinfo {author} {\bibfnamefont
  {H.}~\bibnamefont {Guo}}, \bibinfo {author} {\bibfnamefont {M.}~\bibnamefont
  {P.~M.~Dean}}, \ and\ \bibinfo {author} {\bibfnamefont {X.}~\bibnamefont
  {Liu}},\ }\bibfield  {title} {\enquote {\bibinfo {title} {{Doping evolution
  of the charge excitations and electron correlations in electron-doped
  superconducting $\mathrm{La_{2-\mathit{x}}Ce_{\mathit{x}}CuO_4}$}},}\ }\href
  {\doibase 10.1038/s41535-019-0205-9} {\bibfield  {journal} {\bibinfo
  {journal} {npj Quant. Mater.}\ }\textbf {\bibinfo {volume} {5}},\ \bibinfo
  {pages} {4} (\bibinfo {year} {2020})}\BibitemShut {NoStop}%
\bibitem [{\citenamefont {Singh}\ \emph {et~al.}(2020)\citenamefont {Singh},
  \citenamefont {Huang}, \citenamefont {Lane}, \citenamefont {Li},
  \citenamefont {Okamoto}, \citenamefont {Komiya}, \citenamefont {Markiewicz},
  \citenamefont {Bansil}, \citenamefont {Fujimori}, \citenamefont {Chen},\ and\
  \citenamefont {Huang}}]{SinghArXiV2020}%
  \BibitemOpen
  \bibfield  {author} {\bibinfo {author} {\bibfnamefont {A.}~\bibnamefont
  {Singh}}, \bibinfo {author} {\bibfnamefont {H.~Y.}\ \bibnamefont {Huang}},
  \bibinfo {author} {\bibfnamefont {Christopher}\ \bibnamefont {Lane}},
  \bibinfo {author} {\bibfnamefont {J.~H.}\ \bibnamefont {Li}}, \bibinfo
  {author} {\bibfnamefont {J.}~\bibnamefont {Okamoto}}, \bibinfo {author}
  {\bibfnamefont {S.}~\bibnamefont {Komiya}}, \bibinfo {author} {\bibfnamefont
  {Robert~S.}\ \bibnamefont {Markiewicz}}, \bibinfo {author} {\bibfnamefont
  {Arun}\ \bibnamefont {Bansil}}, \bibinfo {author} {\bibfnamefont
  {A.}~\bibnamefont {Fujimori}}, \bibinfo {author} {\bibfnamefont {C.~T.}\
  \bibnamefont {Chen}}, \ and\ \bibinfo {author} {\bibfnamefont {D.~J.}\
  \bibnamefont {Huang}},\ }\href@noop {} {\enquote {\bibinfo {title} {{Acoustic
  plasmons and conducting carriers in hole-doped cuprate superconductors}},}\ }
  (\bibinfo {year} {2020}),\ \Eprint {http://arxiv.org/abs/2006.13424}
  {arXiv:2006.13424} \BibitemShut {NoStop}%
\bibitem [{\citenamefont {Nag}\ \emph {et~al.}(2020)\citenamefont {Nag},
  \citenamefont {Zhu}, \citenamefont {Bejas}, \citenamefont {Li}, \citenamefont
  {Robarts}, \citenamefont {Yamase}, \citenamefont {Petsch}, \citenamefont
  {Song}, \citenamefont {Eisaki}, \citenamefont {Walters}, \citenamefont
  {Garc\'{\i}a-Fern\'andez}, \citenamefont {Greco}, \citenamefont {Hayden},\
  and\ \citenamefont {Zhou}}]{NagArxiv2020}%
  \BibitemOpen
  \bibfield  {author} {\bibinfo {author} {\bibfnamefont {A.}~\bibnamefont
  {Nag}}, \bibinfo {author} {\bibfnamefont {M.}~\bibnamefont {Zhu}}, \bibinfo
  {author} {\bibfnamefont {M.}~\bibnamefont {Bejas}}, \bibinfo {author}
  {\bibfnamefont {J.}~\bibnamefont {Li}}, \bibinfo {author} {\bibfnamefont
  {H.~C.}\ \bibnamefont {Robarts}}, \bibinfo {author} {\bibfnamefont
  {H.}~\bibnamefont {Yamase}}, \bibinfo {author} {\bibfnamefont {A.~N.}\
  \bibnamefont {Petsch}}, \bibinfo {author} {\bibfnamefont {D.}~\bibnamefont
  {Song}}, \bibinfo {author} {\bibfnamefont {H.}~\bibnamefont {Eisaki}},
  \bibinfo {author} {\bibfnamefont {A.~C.}\ \bibnamefont {Walters}}, \bibinfo
  {author} {\bibfnamefont {M.}~\bibnamefont {Garc\'{\i}a-Fern\'andez}},
  \bibinfo {author} {\bibfnamefont {A.}~\bibnamefont {Greco}}, \bibinfo
  {author} {\bibfnamefont {S.~M.}\ \bibnamefont {Hayden}}, \ and\ \bibinfo
  {author} {\bibfnamefont {K.-J.}\ \bibnamefont {Zhou}},\ }\bibfield  {title}
  {\enquote {\bibinfo {title} {{Detection of Acoustic Plasmons in Hole-Doped
  Lanthanum and Bismuth Cuprate Superconductors Using Resonant Inelastic X-Ray
  Scattering}},}\ }\href {https://doi.org/10.1103/PhysRevLett.125.257002}
  {\bibfield  {journal} {\bibinfo  {journal} {Phys. Rev. Lett.}\ }\textbf
  {\bibinfo {volume} {125}},\ \bibinfo {pages} {257002} (\bibinfo {year}
  {2020})}\BibitemShut {NoStop}%
\bibitem [{\citenamefont {Wakimoto}\ \emph {et~al.}(2004)\citenamefont
  {Wakimoto}, \citenamefont {Zhang}, \citenamefont {Yamada}, \citenamefont
  {Swainson}, \citenamefont {Kim},\ and\ \citenamefont
  {Birgeneau}}]{WakimotoPhysRevLett2004}%
  \BibitemOpen
  \bibfield  {author} {\bibinfo {author} {\bibfnamefont {S.}~\bibnamefont
  {Wakimoto}}, \bibinfo {author} {\bibfnamefont {H.}~\bibnamefont {Zhang}},
  \bibinfo {author} {\bibfnamefont {K.}~\bibnamefont {Yamada}}, \bibinfo
  {author} {\bibfnamefont {I.}~\bibnamefont {Swainson}}, \bibinfo {author}
  {\bibfnamefont {Hyunkyung}\ \bibnamefont {Kim}}, \ and\ \bibinfo {author}
  {\bibfnamefont {R.~J.}\ \bibnamefont {Birgeneau}},\ }\bibfield  {title}
  {\enquote {\bibinfo {title} {{Direct Relation between the Low-Energy Spin
  Excitations and Superconductivity of Overdoped High-${T}_{c}$
  Superconductors}},}\ }\href {\doibase 10.1103/PhysRevLett.92.217004}
  {\bibfield  {journal} {\bibinfo  {journal} {Phys. Rev. Lett.}\ }\textbf
  {\bibinfo {volume} {92}},\ \bibinfo {pages} {217004} (\bibinfo {year}
  {2004})}\BibitemShut {NoStop}%
\bibitem [{\citenamefont {Dahm}\ \emph {et~al.}(2009)\citenamefont {Dahm},
  \citenamefont {Hinkov}, \citenamefont {Borisenko}, \citenamefont {Kordyuk},
  \citenamefont {Zabolotnyy}, \citenamefont {Fink}, \citenamefont
  {B\"{u}chner}, \citenamefont {Scalapino}, \citenamefont {Hanke},\ and\
  \citenamefont {Keimer}}]{DahmNatPhys2009}%
  \BibitemOpen
  \bibfield  {author} {\bibinfo {author} {\bibfnamefont {T.}~\bibnamefont
  {Dahm}}, \bibinfo {author} {\bibfnamefont {V.}~\bibnamefont {Hinkov}},
  \bibinfo {author} {\bibfnamefont {S.~V.}\ \bibnamefont {Borisenko}}, \bibinfo
  {author} {\bibfnamefont {A.~A.}\ \bibnamefont {Kordyuk}}, \bibinfo {author}
  {\bibfnamefont {V.~B.}\ \bibnamefont {Zabolotnyy}}, \bibinfo {author}
  {\bibfnamefont {J.}~\bibnamefont {Fink}}, \bibinfo {author} {\bibfnamefont
  {B.}~\bibnamefont {B\"{u}chner}}, \bibinfo {author} {\bibfnamefont {D.~J.}\
  \bibnamefont {Scalapino}}, \bibinfo {author} {\bibfnamefont {W.}~\bibnamefont
  {Hanke}}, \ and\ \bibinfo {author} {\bibfnamefont {B.}~\bibnamefont
  {Keimer}},\ }\bibfield  {title} {\enquote {\bibinfo {title} {{Strength of the
  spin-fluctuation-mediated pairing interaction in a high-temperature
  superconductor}},}\ }\href {\doibase 10.1038/nphys1180} {\bibfield  {journal}
  {\bibinfo  {journal} {Nat. Phys.}\ }\textbf {\bibinfo {volume} {5}},\
  \bibinfo {pages} {217} (\bibinfo {year} {2009})}\BibitemShut {NoStop}%
\bibitem [{\citenamefont {Grilli}\ \emph {et~al.}(1991)\citenamefont {Grilli},
  \citenamefont {Raimondi}, \citenamefont {Castellani}, \citenamefont
  {Di~Castro},\ and\ \citenamefont {Kotliar}}]{GrilliPhysRevLett1991}%
  \BibitemOpen
  \bibfield  {author} {\bibinfo {author} {\bibfnamefont {M.}~\bibnamefont
  {Grilli}}, \bibinfo {author} {\bibfnamefont {R.}~\bibnamefont {Raimondi}},
  \bibinfo {author} {\bibfnamefont {C.}~\bibnamefont {Castellani}}, \bibinfo
  {author} {\bibfnamefont {C.}~\bibnamefont {Di~Castro}}, \ and\ \bibinfo
  {author} {\bibfnamefont {G.}~\bibnamefont {Kotliar}},\ }\bibfield  {title}
  {\enquote {\bibinfo {title} {{Superconductivity, phase separation, and
  charge-transfer instability in the $U$=\ensuremath{\infty} limit of the
  three-band model of the ${\mathrm{CuO}}_{2}$ planes}},}\ }\href {\doibase
  10.1103/PhysRevLett.67.259} {\bibfield  {journal} {\bibinfo  {journal} {Phys.
  Rev. Lett.}\ }\textbf {\bibinfo {volume} {67}},\ \bibinfo {pages} {259}
  (\bibinfo {year} {1991})}\BibitemShut {NoStop}%
\bibitem [{\citenamefont {Perali}\ \emph {et~al.}(1996)\citenamefont {Perali},
  \citenamefont {Castellani}, \citenamefont {Di~Castro},\ and\ \citenamefont
  {Grilli}}]{PeraliPhysRevB1996}%
  \BibitemOpen
  \bibfield  {author} {\bibinfo {author} {\bibfnamefont {A.}~\bibnamefont
  {Perali}}, \bibinfo {author} {\bibfnamefont {C.}~\bibnamefont {Castellani}},
  \bibinfo {author} {\bibfnamefont {C.}~\bibnamefont {Di~Castro}}, \ and\
  \bibinfo {author} {\bibfnamefont {M.}~\bibnamefont {Grilli}},\ }\bibfield
  {title} {\enquote {\bibinfo {title} {{$d$-wave superconductivity near charge
  instabilities}},}\ }\href {\doibase 10.1103/PhysRevB.54.16216} {\bibfield
  {journal} {\bibinfo  {journal} {Phys. Rev. B}\ }\textbf {\bibinfo {volume}
  {54}},\ \bibinfo {pages} {16216} (\bibinfo {year} {1996})}\BibitemShut
  {NoStop}%
\bibitem [{\citenamefont {Spa{\l}ek}\ \emph {et~al.}(2017)\citenamefont
  {Spa{\l}ek}, \citenamefont {Zegrodnik},\ and\ \citenamefont
  {Kaczmarczyk}}]{SpalekPhysRevB2017}%
  \BibitemOpen
  \bibfield  {author} {\bibinfo {author} {\bibfnamefont {J.}~\bibnamefont
  {Spa{\l}ek}}, \bibinfo {author} {\bibfnamefont {M.}~\bibnamefont
  {Zegrodnik}}, \ and\ \bibinfo {author} {\bibfnamefont {J.}~\bibnamefont
  {Kaczmarczyk}},\ }\bibfield  {title} {\enquote {\bibinfo {title} {{Universal
  properties of high-temperature superconductors from real-space pairing:
  $t$-$J$-$U$ model and its quantitative comparison with experiment}},}\ }\href
  {http://dx.doi.org/10.1103/physrevb.95.024506} {\bibfield  {journal}
  {\bibinfo  {journal} {Phys. Rev. B}\ }\textbf {\bibinfo {volume} {95}},\
  \bibinfo {pages} {024506} (\bibinfo {year} {2017})}\BibitemShut {NoStop}%
\bibitem [{\citenamefont {Greco}\ \emph {et~al.}(2016)\citenamefont {Greco},
  \citenamefont {Yamase},\ and\ \citenamefont {Bejas}}]{GrecoPhysRevB2016}%
  \BibitemOpen
  \bibfield  {author} {\bibinfo {author} {\bibfnamefont {A.}~\bibnamefont
  {Greco}}, \bibinfo {author} {\bibfnamefont {H.}~\bibnamefont {Yamase}}, \
  and\ \bibinfo {author} {\bibfnamefont {M.}~\bibnamefont {Bejas}},\ }\bibfield
   {title} {\enquote {\bibinfo {title} {{Plasmon excitations in layered
  high-$T_c$ cuprates}},}\ }\href
  {http://dx.doi.org/10.1103/physrevb.94.075139} {\bibfield  {journal}
  {\bibinfo  {journal} {Phys. Rev. B}\ }\textbf {\bibinfo {volume} {94}},\
  \bibinfo {pages} {075139} (\bibinfo {year} {2016})}\BibitemShut {NoStop}%
\bibitem [{\citenamefont {Greco}\ \emph {et~al.}(2017)\citenamefont {Greco},
  \citenamefont {Yamase},\ and\ \citenamefont {Bejas}}]{GrecoJPSJ2017}%
  \BibitemOpen
  \bibfield  {author} {\bibinfo {author} {\bibfnamefont {A.}~\bibnamefont
  {Greco}}, \bibinfo {author} {\bibfnamefont {H.}~\bibnamefont {Yamase}}, \
  and\ \bibinfo {author} {\bibfnamefont {M.}~\bibnamefont {Bejas}},\ }\bibfield
   {title} {\enquote {\bibinfo {title} {{Charge-Density-Excitation Spectrum in
  the $t$-$t^\prime$-$J$-$V$ Model}},}\ }\href
  {https://doi.org/10.7566/JPSJ.86.034706} {\bibfield  {journal} {\bibinfo
  {journal} {J. Phys. Soc. Japan}\ }\textbf {\bibinfo {volume} {86}},\ \bibinfo
  {pages} {034706} (\bibinfo {year} {2017})}\BibitemShut {NoStop}%
\bibitem [{\citenamefont {Greco}\ \emph {et~al.}(2020)\citenamefont {Greco},
  \citenamefont {Yamase},\ and\ \citenamefont {Bejas}}]{GrecoPhysRevB2020}%
  \BibitemOpen
  \bibfield  {author} {\bibinfo {author} {\bibfnamefont {A.}~\bibnamefont
  {Greco}}, \bibinfo {author} {\bibfnamefont {H.}~\bibnamefont {Yamase}}, \
  and\ \bibinfo {author} {\bibfnamefont {M.}~\bibnamefont {Bejas}},\ }\bibfield
   {title} {\enquote {\bibinfo {title} {{Close inspection of plasmon
  excitations in cuprate superconductors}},}\ }\href
  {https://doi.org/10.1103/PhysRevB.102.024509} {\bibfield  {journal} {\bibinfo
   {journal} {Phys. Rev. B}\ }\textbf {\bibinfo {volume} {102}},\ \bibinfo
  {pages} {024509} (\bibinfo {year} {2020})}\BibitemShut {NoStop}%
\bibitem [{\citenamefont {Markiewicz}\ \emph {et~al.}(2008)\citenamefont
  {Markiewicz}, \citenamefont {Hasan},\ and\ \citenamefont
  {Bansil}}]{MarkiewiczPhysRevB2008}%
  \BibitemOpen
  \bibfield  {author} {\bibinfo {author} {\bibfnamefont {R.~S.}\ \bibnamefont
  {Markiewicz}}, \bibinfo {author} {\bibfnamefont {M.~Z.}\ \bibnamefont
  {Hasan}}, \ and\ \bibinfo {author} {\bibfnamefont {A.}~\bibnamefont
  {Bansil}},\ }\bibfield  {title} {\enquote {\bibinfo {title} {{Acoustic
  plasmons and doping evolution of Mott physics in resonant inelastic $x$-ray
  scattering from cuprate superconductors}},}\ }\href {\doibase
  10.1103/PhysRevB.77.094518} {\bibfield  {journal} {\bibinfo  {journal} {Phys.
  Rev. B}\ }\textbf {\bibinfo {volume} {77}},\ \bibinfo {pages} {094518}
  (\bibinfo {year} {2008})}\BibitemShut {NoStop}%
\bibitem [{\citenamefont {Greco}\ \emph {et~al.}(2019)\citenamefont {Greco},
  \citenamefont {Yamase},\ and\ \citenamefont {Bejas}}]{GrecoCommunPhys2019}%
  \BibitemOpen
  \bibfield  {author} {\bibinfo {author} {\bibfnamefont {A.}~\bibnamefont
  {Greco}}, \bibinfo {author} {\bibfnamefont {H.}~\bibnamefont {Yamase}}, \
  and\ \bibinfo {author} {\bibfnamefont {M.}~\bibnamefont {Bejas}},\ }\bibfield
   {title} {\enquote {\bibinfo {title} {{Origin of high-energy charge
  excitations observed by resonant inelastic $X$-ray scattering in cuprate
  superconductors}},}\ }\href {\doibase 10.1038/s42005-018-0099-z} {\bibfield
  {journal} {\bibinfo  {journal} {Commun. Phys.}\ }\textbf {\bibinfo {volume}
  {2}},\ \bibinfo {pages} {3} (\bibinfo {year} {2019})}\BibitemShut {NoStop}%
\bibitem [{\citenamefont {Foussats}\ and\ \citenamefont
  {Greco}(2002)}]{FoussatsPhysRevB2002}%
  \BibitemOpen
  \bibfield  {author} {\bibinfo {author} {\bibfnamefont {A.}~\bibnamefont
  {Foussats}}\ and\ \bibinfo {author} {\bibfnamefont {A.}~\bibnamefont
  {Greco}},\ }\bibfield  {title} {\enquote {\bibinfo {title} {{Large-$N$
  expansion based on the Hubbard operator path integral representation and its
  application to the $t$-$J$ model}},}\ }\href {\doibase
  10.1103/PhysRevB.65.195107} {\bibfield  {journal} {\bibinfo  {journal} {Phys.
  Rev. B}\ }\textbf {\bibinfo {volume} {65}},\ \bibinfo {pages} {195107}
  (\bibinfo {year} {2002})}\BibitemShut {NoStop}%
\bibitem [{\citenamefont {Foussats}\ and\ \citenamefont
  {Greco}(2004)}]{FoussatsPhysRevB2004}%
  \BibitemOpen
  \bibfield  {author} {\bibinfo {author} {\bibfnamefont {A.}~\bibnamefont
  {Foussats}}\ and\ \bibinfo {author} {\bibfnamefont {A.}~\bibnamefont
  {Greco}},\ }\bibfield  {title} {\enquote {\bibinfo {title} {{Large-$N$
  expansion based on the Hubbard operator path integral representation and its
  application to the $t$-$J$ model. II. The case for finite $J$}},}\ }\href
  {\doibase 10.1103/physrevb.70.205123} {\bibfield  {journal} {\bibinfo
  {journal} {Phys. Rev. B}\ }\textbf {\bibinfo {volume} {70}},\ \bibinfo
  {pages} {205123} (\bibinfo {year} {2004})}\BibitemShut {NoStop}%
\bibitem [{\citenamefont {Zhang}\ \emph {et~al.}(2020)\citenamefont {Zhang},
  \citenamefont {Wu}, \citenamefont {Kong}, \citenamefont {Bai},\ and\
  \citenamefont {Xu}}]{ZhangJPCM2020}%
  \BibitemOpen
  \bibfield  {author} {\bibinfo {author} {\bibfnamefont {H.-Y.}\ \bibnamefont
  {Zhang}}, \bibinfo {author} {\bibfnamefont {X.-Q.}\ \bibnamefont {Wu}},
  \bibinfo {author} {\bibfnamefont {F.-J.}\ \bibnamefont {Kong}}, \bibinfo
  {author} {\bibfnamefont {Y.-J.}\ \bibnamefont {Bai}}, \ and\ \bibinfo
  {author} {\bibfnamefont {N.}~\bibnamefont {Xu}},\ }\bibfield  {title}
  {\enquote {\bibinfo {title} {{Doping evolution of the magnetic excitations in
  the monolayer $\mathrm{CuO_2}$}},}\ }\href {\doibase
  10.1088/1361-648x/ab99ec} {\bibfield  {journal} {\bibinfo  {journal} {J.
  Phys.: Condens. Matter}\ }\textbf {\bibinfo {volume} {32}},\ \bibinfo {pages}
  {415603} (\bibinfo {year} {2020})}\BibitemShut {NoStop}%
\bibitem [{\citenamefont {Fidrysiak}\ and\ \citenamefont
  {Spa{\l}ek}(2020)}]{FidrysiakPhysRevB2020}%
  \BibitemOpen
  \bibfield  {author} {\bibinfo {author} {\bibfnamefont {M.}~\bibnamefont
  {Fidrysiak}}\ and\ \bibinfo {author} {\bibfnamefont {J.}~\bibnamefont
  {Spa{\l}ek}},\ }\bibfield  {title} {\enquote {\bibinfo {title} {{Robust spin
  and charge excitations throughout the high-$T_c$ cuprate phase diagram from
  incipient Mottness}},}\ }\href
  {http://dx.doi.org/10.1103/physrevb.102.014505} {\bibfield  {journal}
  {\bibinfo  {journal} {Phys. Rev. B}\ }\textbf {\bibinfo {volume} {102}},\
  \bibinfo {pages} {014505} (\bibinfo {year} {2020})}\BibitemShut {NoStop}%
\bibitem [{\citenamefont {Fidrysiak}\ and\ \citenamefont
  {Spa\l{}ek}(2021)}]{FidrysiakArXiv2020}%
  \BibitemOpen
  \bibfield  {author} {\bibinfo {author} {\bibfnamefont {M.}~\bibnamefont
  {Fidrysiak}}\ and\ \bibinfo {author} {\bibfnamefont {J.}~\bibnamefont
  {Spa\l{}ek}},\ }\bibfield  {title} {\enquote {\bibinfo {title} {{Universal
  collective modes from strong electronic correlations: Modified
  $1/{\mathcal{N}}_{f}$ theory with application to high-${T}_{c}$ cuprates}},}\
  }\href {\doibase 10.1103/PhysRevB.103.165111} {\bibfield  {journal} {\bibinfo
   {journal} {Phys. Rev. B}\ }\textbf {\bibinfo {volume} {103}},\ \bibinfo
  {pages} {165111} (\bibinfo {year} {2021})}\BibitemShut {NoStop}%
\bibitem [{\citenamefont {Zegrodnik}\ and\ \citenamefont
  {Spa\l{}ek}(2017{\natexlab{a}})}]{ZegrodnikPhysRevB2017_2}%
  \BibitemOpen
  \bibfield  {author} {\bibinfo {author} {\bibfnamefont {M.}~\bibnamefont
  {Zegrodnik}}\ and\ \bibinfo {author} {\bibfnamefont {J.}~\bibnamefont
  {Spa\l{}ek}},\ }\bibfield  {title} {\enquote {\bibinfo {title} {{Universal
  properties of high-temperature superconductors from real-space pairing: Role
  of correlated hopping and intersite Coulomb interaction within the
  $t$-$J$-$U$ model}},}\ }\href {https://doi.org/10.1103/PhysRevB.96.054511}
  {\bibfield  {journal} {\bibinfo  {journal} {Phys. Rev. B}\ }\textbf {\bibinfo
  {volume} {96}},\ \bibinfo {pages} {054511} (\bibinfo {year}
  {2017}{\natexlab{a}})}\BibitemShut {NoStop}%
\bibitem [{\citenamefont {Zegrodnik}\ and\ \citenamefont
  {Spa\l{}ek}(2017{\natexlab{b}})}]{ZegrodnikPhysRevB2017_3}%
  \BibitemOpen
  \bibfield  {author} {\bibinfo {author} {\bibfnamefont {M.}~\bibnamefont
  {Zegrodnik}}\ and\ \bibinfo {author} {\bibfnamefont {J.}~\bibnamefont
  {Spa\l{}ek}},\ }\bibfield  {title} {\enquote {\bibinfo {title} {{Effect of
  interlayer processes on the superconducting state within the $t$-$J$-$U$
  model: Full Gutzwiller wave-function solution and relation to experiment}},}\
  }\href {\doibase 10.1103/PhysRevB.95.024507} {\bibfield  {journal} {\bibinfo
  {journal} {Phys. Rev. B}\ }\textbf {\bibinfo {volume} {95}},\ \bibinfo
  {pages} {024507} (\bibinfo {year} {2017}{\natexlab{b}})}\BibitemShut
  {NoStop}%
\bibitem [{\citenamefont {Fidrysiak}\ \emph {et~al.}(2018)\citenamefont
  {Fidrysiak}, \citenamefont {Zegrodnik},\ and\ \citenamefont
  {Spa{\l}ek}}]{FidrysiakJPhysCondensMatter2018}%
  \BibitemOpen
  \bibfield  {author} {\bibinfo {author} {\bibfnamefont {M.}~\bibnamefont
  {Fidrysiak}}, \bibinfo {author} {\bibfnamefont {M.}~\bibnamefont
  {Zegrodnik}}, \ and\ \bibinfo {author} {\bibfnamefont {J.}~\bibnamefont
  {Spa{\l}ek}},\ }\bibfield  {title} {\enquote {\bibinfo {title} {{Realistic
  estimates of superconducting properties for the cuprates: reciprocal-space
  diagrammatic expansion combined with variational approach}},}\ }\href
  {http://dx.doi.org/10.1088/1361-648X/aae6fb} {\bibfield  {journal} {\bibinfo
  {journal} {J. Phys.: Condens. Matter}\ }\textbf {\bibinfo {volume} {30}},\
  \bibinfo {pages} {475602} (\bibinfo {year} {2018})}\BibitemShut {NoStop}%
\bibitem [{\citenamefont {Zegrodnik}\ and\ \citenamefont
  {Spa{\l}ek}(2018)}]{ZegrodnikPhysRevB2018}%
  \BibitemOpen
  \bibfield  {author} {\bibinfo {author} {\bibfnamefont {M.}~\bibnamefont
  {Zegrodnik}}\ and\ \bibinfo {author} {\bibfnamefont {J.}~\bibnamefont
  {Spa{\l}ek}},\ }\bibfield  {title} {\enquote {\bibinfo {title}
  {{Incorporation of charge- and pair-density-wave states into the one-band
  model of $d$-wave superconductivity}},}\ }\href {\doibase
  10.1103/physrevb.98.155144} {\bibfield  {journal} {\bibinfo  {journal} {Phys.
  Rev. B}\ }\textbf {\bibinfo {volume} {98}},\ \bibinfo {pages} {155144}
  (\bibinfo {year} {2018})}\BibitemShut {NoStop}%
\bibitem [{\citenamefont {Zegrodnik}\ \emph {et~al.}(2020)\citenamefont
  {Zegrodnik}, \citenamefont {Biborski}, \citenamefont {Fidrysiak},\ and\
  \citenamefont {Spa{\l}ek}}]{ZegrodnikJPCM2021}%
  \BibitemOpen
  \bibfield  {author} {\bibinfo {author} {\bibfnamefont {M.}~\bibnamefont
  {Zegrodnik}}, \bibinfo {author} {\bibfnamefont {A.}~\bibnamefont {Biborski}},
  \bibinfo {author} {\bibfnamefont {M.}~\bibnamefont {Fidrysiak}}, \ and\
  \bibinfo {author} {\bibfnamefont {J.}~\bibnamefont {Spa{\l}ek}},\ }\href@noop
  {} {\enquote {\bibinfo {title} {{Superconductivity in the three-band model of
  cuprates: nodal direction characteristics and influence of intersite
  interactions}},}\ } (\bibinfo {year} {2020}),\ \bibinfo {note} {{J. Phys.:
  Condens. Matter (in press)}},\ \Eprint {http://arxiv.org/abs/2009.04922}
  {arXiv:2009.04922} \BibitemShut {NoStop}%
\bibitem [{\citenamefont {Nilsson}\ \emph {et~al.}(2019)\citenamefont
  {Nilsson}, \citenamefont {Karlsson},\ and\ \citenamefont
  {Aryasetiawan}}]{NilssonPhysRevB2019}%
  \BibitemOpen
  \bibfield  {author} {\bibinfo {author} {\bibfnamefont {F.}~\bibnamefont
  {Nilsson}}, \bibinfo {author} {\bibfnamefont {K.}~\bibnamefont {Karlsson}}, \
  and\ \bibinfo {author} {\bibfnamefont {F.}~\bibnamefont {Aryasetiawan}},\
  }\bibfield  {title} {\enquote {\bibinfo {title} {{Dynamically screened
  Coulomb interaction in the parent compounds of hole-doped cuprates: Trends
  and exceptions}},}\ }\href {\doibase 10.1103/PhysRevB.99.075135} {\bibfield
  {journal} {\bibinfo  {journal} {Phys. Rev. B}\ }\textbf {\bibinfo {volume}
  {99}},\ \bibinfo {pages} {075135} (\bibinfo {year} {2019})}\BibitemShut
  {NoStop}%
\bibitem [{\citenamefont {Becca}\ \emph {et~al.}(1996)\citenamefont {Becca},
  \citenamefont {Tarquini}, \citenamefont {Grilli},\ and\ \citenamefont
  {Di~Castro}}]{BeccaPhysRevB1996}%
  \BibitemOpen
  \bibfield  {author} {\bibinfo {author} {\bibfnamefont {F.}~\bibnamefont
  {Becca}}, \bibinfo {author} {\bibfnamefont {M.}~\bibnamefont {Tarquini}},
  \bibinfo {author} {\bibfnamefont {M.}~\bibnamefont {Grilli}}, \ and\ \bibinfo
  {author} {\bibfnamefont {C.}~\bibnamefont {Di~Castro}},\ }\bibfield  {title}
  {\enquote {\bibinfo {title} {{Charge-density waves and superconductivity as
  an alternative to phase separation in the infinite-U Hubbard-Holstein
  model}},}\ }\href {\doibase 10.1103/PhysRevB.54.12443} {\bibfield  {journal}
  {\bibinfo  {journal} {Phys. Rev. B}\ }\textbf {\bibinfo {volume} {54}},\
  \bibinfo {pages} {12443} (\bibinfo {year} {1996})}\BibitemShut {NoStop}%
\bibitem [{\citenamefont {van Heumen}\ \emph {et~al.}(2009)\citenamefont {van
  Heumen}, \citenamefont {Meevasana}, \citenamefont {Kuzmenko}, \citenamefont
  {Eisaki},\ and\ \citenamefont {van~der Marel}}]{vanHeumenN2009}%
  \BibitemOpen
  \bibfield  {author} {\bibinfo {author} {\bibfnamefont {E.}~\bibnamefont {van
  Heumen}}, \bibinfo {author} {\bibfnamefont {W.}~\bibnamefont {Meevasana}},
  \bibinfo {author} {\bibfnamefont {A.~B.}\ \bibnamefont {Kuzmenko}}, \bibinfo
  {author} {\bibfnamefont {H.}~\bibnamefont {Eisaki}}, \ and\ \bibinfo {author}
  {\bibfnamefont {D.}~\bibnamefont {van~der Marel}},\ }\bibfield  {title}
  {\enquote {\bibinfo {title} {{Doping-dependent optical properties of
  Bi2201}},}\ }\href {\doibase 10.1088/1367-2630/11/5/055067} {\bibfield
  {journal} {\bibinfo  {journal} {New J. Phys.}\ }\textbf {\bibinfo {volume}
  {11}},\ \bibinfo {pages} {055067} (\bibinfo {year} {2009})}\BibitemShut
  {NoStop}%
\bibitem [{\citenamefont {Hirayama}\ \emph {et~al.}(2018)\citenamefont
  {Hirayama}, \citenamefont {Yamaji}, \citenamefont {Misawa},\ and\
  \citenamefont {Imada}}]{HirayamaPhysRevB2018}%
  \BibitemOpen
  \bibfield  {author} {\bibinfo {author} {\bibfnamefont {M.}~\bibnamefont
  {Hirayama}}, \bibinfo {author} {\bibfnamefont {Y.}~\bibnamefont {Yamaji}},
  \bibinfo {author} {\bibfnamefont {T.}~\bibnamefont {Misawa}}, \ and\ \bibinfo
  {author} {\bibfnamefont {M.}~\bibnamefont {Imada}},\ }\bibfield  {title}
  {\enquote {\bibinfo {title} {{Ab initio effective Hamiltonians for cuprate
  superconductors}},}\ }\href {\doibase 10.1103/physrevb.98.134501} {\bibfield
  {journal} {\bibinfo  {journal} {Phys. Rev. B}\ }\textbf {\bibinfo {volume}
  {98}},\ \bibinfo {pages} {134501} (\bibinfo {year} {2018})}\BibitemShut
  {NoStop}%
\bibitem [{\citenamefont {Igoshev}\ \emph {et~al.}(2015)\citenamefont
  {Igoshev}, \citenamefont {Timirgazin}, \citenamefont {Gilmutdinov},
  \citenamefont {Arzhnikov},\ and\ \citenamefont
  {Yu~Irkhin}}]{IgoshevJPCM2015}%
  \BibitemOpen
  \bibfield  {author} {\bibinfo {author} {\bibfnamefont {P.~A.}\ \bibnamefont
  {Igoshev}}, \bibinfo {author} {\bibfnamefont {M.~A.}\ \bibnamefont
  {Timirgazin}}, \bibinfo {author} {\bibfnamefont {V.~F.}\ \bibnamefont
  {Gilmutdinov}}, \bibinfo {author} {\bibfnamefont {A.~K.}\ \bibnamefont
  {Arzhnikov}}, \ and\ \bibinfo {author} {\bibfnamefont {V.}~\bibnamefont
  {Yu~Irkhin}},\ }\bibfield  {title} {\enquote {\bibinfo {title} {{Spiral
  magnetism in the single-band Hubbard model: the Hartree-Fock and slave-boson
  approaches}},}\ }\href {\doibase 10.1088/0953-8984/27/44/446002} {\bibfield
  {journal} {\bibinfo  {journal} {J. Phys.: Condens. Matter}\ }\textbf
  {\bibinfo {volume} {27}},\ \bibinfo {pages} {446002} (\bibinfo {year}
  {2015})}\BibitemShut {NoStop}%
\bibitem [{\citenamefont {J\k{e}drak}\ and\ \citenamefont
  {Spa{\l}ek}(2011)}]{JedrakPhsRevB2011}%
  \BibitemOpen
  \bibfield  {author} {\bibinfo {author} {\bibfnamefont {J.}~\bibnamefont
  {J\k{e}drak}}\ and\ \bibinfo {author} {\bibfnamefont {J.}~\bibnamefont
  {Spa{\l}ek}},\ }\bibfield  {title} {\enquote {\bibinfo {title} {{Renormalized
  mean-field $t$-$J$ model of high-$T_c$ superconductivity: Comparison to
  experiment}},}\ }\href {http://dx.doi.org/10.1103/physrevb.83.104512}
  {\bibfield  {journal} {\bibinfo  {journal} {Phys. Rev. B}\ }\textbf {\bibinfo
  {volume} {83}},\ \bibinfo {pages} {104512} (\bibinfo {year}
  {2011})}\BibitemShut {NoStop}%
\bibitem [{\citenamefont {Lamsal}\ and\ \citenamefont
  {Montfrooij}(2016)}]{LamsalPhysRevB2016}%
  \BibitemOpen
  \bibfield  {author} {\bibinfo {author} {\bibfnamefont {J.}~\bibnamefont
  {Lamsal}}\ and\ \bibinfo {author} {\bibfnamefont {W.}~\bibnamefont
  {Montfrooij}},\ }\bibfield  {title} {\enquote {\bibinfo {title} {{Extracting
  paramagnon excitations from resonant inelastic $x$-ray scattering
  experiments}},}\ }\href {http://dx.doi.org/10.1103/physrevb.93.214513}
  {\bibfield  {journal} {\bibinfo  {journal} {Phys. Rev. B}\ }\textbf {\bibinfo
  {volume} {93}},\ \bibinfo {pages} {214513} (\bibinfo {year}
  {2016})}\BibitemShut {NoStop}%
\bibitem [{\citenamefont {Kovaleva}\ \emph {et~al.}(2004)\citenamefont
  {Kovaleva}, \citenamefont {Boris}, \citenamefont {Holden}, \citenamefont
  {Ulrich}, \citenamefont {Liang}, \citenamefont {Lin}, \citenamefont {Keimer},
  \citenamefont {Bernhard}, \citenamefont {Tallon}, \citenamefont {Munzar},\
  and\ \citenamefont {Stoneham}}]{KovalevaPhysRevB2004}%
  \BibitemOpen
  \bibfield  {author} {\bibinfo {author} {\bibfnamefont {N.~N.}\ \bibnamefont
  {Kovaleva}}, \bibinfo {author} {\bibfnamefont {A.~V.}\ \bibnamefont {Boris}},
  \bibinfo {author} {\bibfnamefont {T.}~\bibnamefont {Holden}}, \bibinfo
  {author} {\bibfnamefont {C.}~\bibnamefont {Ulrich}}, \bibinfo {author}
  {\bibfnamefont {B.}~\bibnamefont {Liang}}, \bibinfo {author} {\bibfnamefont
  {C.~T.}\ \bibnamefont {Lin}}, \bibinfo {author} {\bibfnamefont
  {B.}~\bibnamefont {Keimer}}, \bibinfo {author} {\bibfnamefont
  {C.}~\bibnamefont {Bernhard}}, \bibinfo {author} {\bibfnamefont {J.~L.}\
  \bibnamefont {Tallon}}, \bibinfo {author} {\bibfnamefont {D.}~\bibnamefont
  {Munzar}}, \ and\ \bibinfo {author} {\bibfnamefont {A.~M.}\ \bibnamefont
  {Stoneham}},\ }\bibfield  {title} {\enquote {\bibinfo {title} {{$c$-axis
  lattice dynamics in Bi-based cuprate superconductors}},}\ }\href {\doibase
  10.1103/PhysRevB.69.054511} {\bibfield  {journal} {\bibinfo  {journal} {Phys.
  Rev. B}\ }\textbf {\bibinfo {volume} {69}},\ \bibinfo {pages} {054511}
  (\bibinfo {year} {2004})}\BibitemShut {NoStop}%
\bibitem [{\citenamefont {Spa{\l}ek}(1988)}]{SpalekPhysRevB1988}%
  \BibitemOpen
  \bibfield  {author} {\bibinfo {author} {\bibfnamefont {J.}~\bibnamefont
  {Spa{\l}ek}},\ }\bibfield  {title} {\enquote {\bibinfo {title} {{Effect of
  pair hopping and magnitude of intra-atomic interaction on exchange-mediated
  superconductivity}},}\ }\href {http://dx.doi.org/10.1103/physrevb.37.533}
  {\bibfield  {journal} {\bibinfo  {journal} {Phys. Rev. B}\ }\textbf {\bibinfo
  {volume} {37}},\ \bibinfo {pages} {533} (\bibinfo {year} {1988})}\BibitemShut
  {NoStop}%
\bibitem [{\citenamefont {Abram}\ \emph {et~al.}(2017)\citenamefont {Abram},
  \citenamefont {Zegrodnik},\ and\ \citenamefont
  {Spa{\l}ek}}]{AbramJPhysCondensMatter2017}%
  \BibitemOpen
  \bibfield  {author} {\bibinfo {author} {\bibfnamefont {M.}~\bibnamefont
  {Abram}}, \bibinfo {author} {\bibfnamefont {M.}~\bibnamefont {Zegrodnik}}, \
  and\ \bibinfo {author} {\bibfnamefont {J.}~\bibnamefont {Spa{\l}ek}},\
  }\bibfield  {title} {\enquote {\bibinfo {title} {{Antiferromagnetism, charge
  density wave, andd-wave superconductivity in the extended $t$-$J$-$U$ model:
  role of intersite Coulomb interaction and a critical overview of renormalized
  mean field theory}},}\ }\href {http://dx.doi.org/10.1088/1361-648X/aa7a21}
  {\bibfield  {journal} {\bibinfo  {journal} {J. Phys.: Condens. Matter}\
  }\textbf {\bibinfo {volume} {29}},\ \bibinfo {pages} {365602} (\bibinfo
  {year} {2017})}\BibitemShut {NoStop}%
\end{thebibliography}
\end{document}